\documentclass[
    aps,
    prapplied,
    reprint,
    floatfix,
    nofootinbib
]{revtex4-2}

\usepackage[T1]{fontenc}
\usepackage[utf8]{inputenc}

\usepackage{amsmath,amssymb,amsfonts}
\usepackage{graphicx}
\usepackage{physics}
\usepackage{booktabs}
\usepackage{tikz}
\usetikzlibrary{arrows.meta,positioning}
\usepackage{hyperref}   

\newcommand{\grel}{\gamma_{\mathrm{rel}}}
\newcommand{\gdeph}{\gamma_\phi}
\newcommand{\gx}{\gamma_x}
\newcommand{\gz}{\gamma_z}
\newcommand{\Ox}{\Omega_x}
\newcommand{\Sth}{S_{\mathrm{th}}}

\newcommand{\etatrue}{\eta_{\mathrm{true}}}
\newcommand{\etaa}{\eta_{\mathrm{a}}}

\newcommand{\Fgate}{F_{\mathrm{gate}}}
\newcommand{\Fmm}{F_{\mathrm{mm}}}

\begin{document}

\title{Coherence-gated quantum devices via weak continuous measurement}

\author{Priyank Singh}
\email{priyank@iitmandi.ac.in}
\affiliation{Centre for Quantum Science and Technologies (CQST),
Indian Institute of Technology Mandi, Himachal Pradesh 175005, India}

\date{\today}

\begin{abstract}
Single-photon routers in cavity and circuit QED direct photons by
the qubit's energy eigenstate--a projective decision that destroys
coherence. We propose a different primitive:
\emph{coherence-gated routing}, where the decision depends on the
magnitude of the qubit's quantum coherence, estimated in real time
from simultaneous weak measurements of $\sigma_x$ and $\sigma_z$.
A photon is accepted if the coherence score
$S(T) = \sqrt{\langle\sigma_x\rangle_c^2 +
\langle\sigma_y\rangle_c^2}$, extracted from the conditional density
matrix via the stochastic master equation, exceeds a tunable
threshold~$\Sth$.

Certifying coherence at the moment of emission enables two
applications that conventional heralded sources cannot provide:
(i)~a quantum random number generator with min-entropy bounded by
Bloch-sphere geometry, $H_\infty \geq
-\log_2\!\bigl(\frac{1+\sqrt{1-\Sth^2}}{2}\bigr)$, and (ii)~a
phase-tracked photon source whose two-node coherence certification
bounds the matter--matter entanglement fidelity after Bell-state
measurement.

The estimator is itself a security primitive. Benchmarking seven
configurations, we find that deliberately underestimating detector
efficiency ($\etaa < \etatrue$) both stabilizes the numerics and
suppresses overcertification. We trace this analytically via a
purity-production monotonicity argument, identify a geometric loophole
that amplifies purity undercertification into coherence
overcertification by more than an order of magnitude (${\sim}40\times$),
and establish two complementary tail bounds: an
Ornstein--Uhlenbeck comparison--giving $4.5\%$ raw overcertification
(empirical $3.7\%$ from $10^6$ trajectories) and sub-percent failure
probabilities for the joint acceptance--failure security statements--and
an exponential supermartingale establishing the structural
existence of an exponential tail bound.
\end{abstract}


\maketitle

\section{Introduction}
\label{sec:intro}

Single-photon routing--directing individual photons to specified
output ports--is a fundamental primitive in quantum
networks~\cite{Kimble2008_QuantumInternet,Wehner2018_QuantumInternet}.
Existing schemes based on cavity quantum electrodynamics
(QED)~\cite{Hoi2011_Router,Shomroni2014_AllOpticalRouter,
Aoki2009_EfficientRouter} and circuit
QED~\cite{Zhou2013_DualChannel,Sliwa2015_Circulator} typically condition
the routing decision on the qubit's energy eigenstate: the photon is
directed one way if the qubit is in $\ket{g}$ (ground state) and another if it is
in~$\ket{e}$ (excited state). This state-based approach requires projective measurement,
which destroys any pre-existing coherence and limits the router to binary
classical control.

In this work we introduce a fundamentally different paradigm:
\emph{coherence-gated routing}, where the routing decision is conditioned
not on which eigenstate the qubit occupies, but on the degree of quantum
coherence--the magnitude of the off-diagonal elements of the conditional
density matrix. This shifts the question from ``what state is the qubit
in?'' to ``how well-defined is the qubit's quantum state?'' and uses that
answer to certify photon quality in a way no post-processing on the
photon alone can achieve.

Coherence is a basis-dependent quantity; here we use the
$\ell_1$-norm of coherence in the $\sigma_z$
eigenbasis~\cite{Baumgratz2014_Coherence,Streltsov2017_Colloquium}--a
choice dictated by a self-consistency requirement between the
measurement axes, the coherence basis, and the Hamiltonian drive,
developed fully in Sec.~\ref{sec:triad}:
\begin{equation}
    S(t) \equiv C_{\ell_1}(\rho_c) = 2|\rho_{01}^{(c)}(t)|
    = \sqrt{\langle\sigma_x\rangle_c^2
           + \langle\sigma_y\rangle_c^2},
    \label{eq:coherence}
\end{equation}
which requires access to the \emph{conditional} state $\rho_c(t)$--the
Bayesian posterior over quantum states given the measurement record up to
time~$t$. Extracting this conditional coherence demands continuous weak
measurement of two non-commuting
observables~\cite{Murch2013_trajectories,
Weber2016_review,HacohenGourgy2016_Simultaneous,
Jordan2006_QubitFeedback},
followed by real-time state estimation via the stochastic master equation
(SME)~\cite{Wiseman2009_Book,Jacobs2014_Book}. This scheme lies at the
intersection of quantum trajectory
theory~\cite{Carmichael1993_Book,Dalibard1992_WaveFunctionApproach},
quantum state
estimation~\cite{VanHandel2005_StabilityFilter,Ralph2011_ContinuousEstimation},
and quantum coherence as a
resource~\cite{Baumgratz2014_Coherence,Streltsov2017_Colloquium}.

The key insight of this paper is that coherence certification at the
moment of photon emission enables applications qualitatively
inaccessible to conventional heralded sources:
\begin{enumerate}
    \item A \emph{coherence-certified quantum random number generator}
    (QRNG) whose min-entropy per bit is bounded by Bloch-sphere
    geometry (Sec.~\ref{sec:qrng}).
    \item A \emph{coherence-certified, phase-tracked photon source}
    for quantum networks, with classical feedforward phase correction
    (Sec.~\ref{sec:qnet}).
\end{enumerate}
We further show how independent coherence certification at two nodes
constrains the achievable entanglement quality under an idealized
Bell-state measurement model, and discuss the assumptions
required (Sec.~\ref{sec:entanglement}).

In both primary applications, the estimator's fidelity to the true
conditional state is not merely a numerical concern--it is a
\emph{security and fidelity requirement}. This elevates the real-time
estimation problem from a methods exercise to an application-critical
component of the device, and motivates the systematic estimator
benchmarking in Sec.~\ref{sec:estimators} and the pointwise
overcertification analysis in Sec.~\ref{sec:pointwise}.

The paper is organized as follows. Section~\ref{sec:protocol} describes
the protocol and theoretical model.
Section~\ref{sec:triad} establishes the measurement--coherence--drive
self-consistency triad.
Section~\ref{sec:qrng} develops the QRNG application, including a
comparison with existing protocols.
Sections~\ref{sec:qnet} and~\ref{sec:entanglement} develop quantum
networking and entanglement distribution.
Section~\ref{sec:certification} analyzes certification integrity and
the geometric loophole from purity to coherence.
Section~\ref{sec:pointwise} develops two complementary pointwise
overcertification bounds and the composable security statement.
Section~\ref{sec:estimators} presents the estimator hierarchy.
Section~\ref{sec:analytical} examines the assumed efficiency,
distinguishing decision accuracy from overcertification.
Section~\ref{sec:discussion} discusses hardware scaling and
experimental prospects. Section~\ref{sec:conclusions} concludes the work.

\section{Protocol and theoretical framework}
\label{sec:protocol}

\begin{figure}[tb]
\centering
\begin{tikzpicture}[x=0.9cm, y=0.9cm, >=Latex,
    box/.style={draw, thick, rounded corners=2pt, minimum height=0.7cm,
                inner sep=3pt, font=\footnotesize},
    circ/.style={draw, thick, circle, minimum size=0.6cm,
                 font=\footnotesize}]
\node[circ, fill=blue!15] (Q) at (0,0) {$Q$};
\node[font=\scriptsize, below=1pt] at (Q.south) {qubit};
\node[box, fill=green!10] (Mx) at (-2.2,1.5) {$I_x(t)$};
\node[box, fill=green!10] (Mz) at (2.2,1.5) {$I_z(t)$};
\draw[->, thick] (Q) -- node[left, font=\tiny] {$\sigma_x$} (Mx);
\draw[->, thick] (Q) -- node[right, font=\tiny] {$\sigma_z$} (Mz);
\node[box, fill=red!10, minimum width=2.5cm] (FPGA) at (0,3)
     {FPGA: $\rho_c(t)$};
\draw[->, thick] (Mx) -- (FPGA);
\draw[->, thick] (Mz) -- (FPGA);
\node[box, fill=orange!10] (Dec) at (0,4.5)
     {$S(T) \gtrless \Sth$};
\draw[->, thick] (FPGA) -- (Dec);
\node[box, fill=yellow!10] (RA) at (-2,6) {Port A};
\node[box, fill=yellow!10] (RB) at (2,6) {Port B};
\draw[->, thick] (Dec) -- node[left, font=\tiny] {$S > \Sth$} (RA);
\draw[->, thick] (Dec) -- node[right, font=\tiny] {$S \le \Sth$} (RB);
\node[circ, fill=purple!10] (S) at (0,6) {$a_S$};
\draw[->, dashed, thick] (S) -- (RA);
\draw[->, dashed, thick] (S) -- (RB);
\end{tikzpicture}
\caption{Architecture of the coherence-gated single-photon router.
Simultaneous weak measurements of $\sigma_x$ and $\sigma_z$ yield
homodyne currents $I_x(t)$ and $I_z(t)$. An FPGA updates the
conditional state $\rho_c(t)$ in real time and computes the coherence
score $S(T)$ at decision time $T$. Based on the comparison
$S(T) \gtrless \Sth$, a routing pulse directs the photon from the
source cavity $a_S$ to port~A (high coherence) or port~B (low
coherence).}
\label{fig:architecture}
\end{figure}
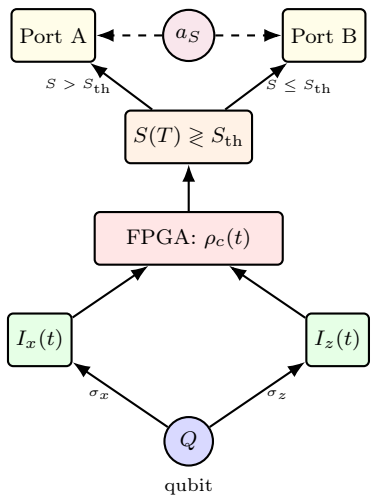

The overall architecture is summarized in
Fig.~\ref{fig:architecture}. 

A driven qubit is monitored
simultaneously along two non-commuting axes ($\sigma_x$ and
$\sigma_z$); an FPGA reconstructs the conditional state in real time
from the resulting homodyne records; and the routing decision is
conditioned on the coherence score extracted from that state. The
remainder of this section develops each component in turn.

\subsection{Qubit Hamiltonian and decoherence}

We consider a qubit with Hamiltonian
\begin{equation}
    H_q = \tfrac{1}{2}\Ox\sigma_x + \tfrac{1}{2}\Delta\sigma_z,
    \label{eq:Hq}
\end{equation}
where $\Ox$ is the transverse Rabi drive strength, $\Delta$ is the
detuning and $\sigma_k$ are the Pauli matrices. The qubit is subject to two intrinsic decoherence channels:
energy relaxation at rate $\grel$ (described by the Lindblad operator
$\sqrt{\grel}\,\sigma_-$, where $\sigma_- = \ket{0}\bra{1}$) and
pure dephasing at rate $\gdeph$ (Lindblad operator
$\sqrt{\gdeph}\,\sigma_z$). We denote the sum of intrinsic
rates as $\gamma_\mathrm{dec} = \grel + \gdeph$.
Throughout we adopt the convention
$\sigma_z = \ket{0}\bra{0} - \ket{1}\bra{1}$, so that $\ket{0}$ is the
relaxation fixed point with $\langle\sigma_z\rangle = +1$; the lowering
operator $\sigma_- = \ket{0}\bra{1}$ accordingly drives the conditional
state toward $z \to +1$, consistent with the Born-rule weight
$P_z = (1\pm z)/2$ of Sec.~\ref{sec:qrng} and the $(1\pm z_i)/2$ pair
weights of Sec.~\ref{sec:entanglement}.
(The exact transverse dephasing contribution from relaxation is
$\grel/2$, not $\grel$; this distinction is absorbed into the
approximate character of $S_\mathrm{typ}$ in Sec.~\ref{sec:analytical}.)

\subsection{Simultaneous weak measurement}

In addition to intrinsic decoherence, the qubit is simultaneously
monitored by two homodyne detection channels that weakly measure
$\sigma_x$ and $\sigma_z$ with measurement rates $\gx$ and $\gz$,
respectively. Each channel produces a continuous photocurrent. In
practice, only a fraction of the outgoing signal is captured by the
detector; this fraction is quantified by the \emph{measurement
efficiency} $\eta_k \in [0,1]$ for channel $k$. The monitored fraction
$\eta_k$ contributes measurement backaction that is captured in the
conditional state update, while the unmonitored fraction $(1-\eta_k)$
acts as additional dephasing. For simplicity we take $\eta_x = \eta_z
\equiv \eta$ throughout.

The homodyne photocurrents are~\cite{Wiseman2009_Book,Jacobs2014_Book}
\begin{equation}
    I_k(t) = \sqrt{4\eta\gamma_k}\,\langle\sigma_k\rangle_c(t)
    + \xi_k(t), \quad k \in \{x,z\},
    \label{eq:currents}
\end{equation}
where $\langle\sigma_k\rangle_c(t) = \mathrm{Tr}[\sigma_k \rho_c(t)]$
is the conditional expectation of $\sigma_k$ and $\xi_k(t)$ is Gaussian
white noise satisfying $\langle \xi_k(t)\xi_{k'}(t')\rangle =
\delta_{kk'}\delta(t-t')$. The factor of~4 arises because the
signal is proportional to
$\langle L_k + L_k^\dagger \rangle = 2\sqrt{\gamma_k}\,
\langle\sigma_k\rangle_c$ for Hermitian $\sigma_k$.
The signal-to-noise ratio of each current
is set by the product $\eta\gamma_k$: larger measurement rate or
higher efficiency produces a stronger signal relative to the shot noise
floor.

\subsection{Stochastic master equation}

The conditional density matrix $\rho_c(t)$ evolves under the stochastic
master equation (SME) in It\=o
form~\cite{Wiseman2009_Book,Jacobs2014_Book}:
\begin{align}
    d\rho_c &= -i[H_q,\rho_c]\,dt \nonumber\\
    &\quad + \grel\,\mathcal{D}[\sigma_-]\rho_c\,dt
    + \gdeph\,\mathcal{D}[\sigma_z]\rho_c\,dt \nonumber\\
    &\quad + \sum_{k=x,z} \gamma_k\,
            \mathcal{D}[\sigma_k]\rho_c\,dt \nonumber\\
    &\quad + \sum_{k=x,z}\sqrt{\eta\gamma_k}\,
            \mathcal{H}[\sigma_k]\rho_c\,dW_k.
    \label{eq:SME}
\end{align}
Here $\mathcal{D}[L]\rho = L\rho L^\dagger - \frac{1}{2}(L^\dagger L
\rho + \rho L^\dagger L)$ is the Lindblad dissipator and
$\mathcal{H}[L]\rho = L\rho + \rho L^\dagger -
\mathrm{Tr}[(L+L^\dagger)\rho]\,\rho$ is the measurement backaction
(innovation) superoperator. The Wiener increments
$dW_k$ satisfy $dW_k\,dW_{k'} = \delta_{kk'}\,dt$ and are related
to the photocurrents by $dW_k = I_k(t)\,dt -
\sqrt{4\eta\gamma_k}\,\langle\sigma_k\rangle_c\,dt$.

The structure of Eq.~\eqref{eq:SME} reflects the standard decomposition
of homodyne monitoring with efficiency $\eta < 1$
(Ref.~\cite{Wiseman2009_Book}, Eq.~4.238). Each measurement channel
couples to the qubit at the full rate $\gamma_k$, contributing
$\gamma_k\,\mathcal{D}[\sigma_k]\rho_c$ to the deterministic
evolution regardless of detector efficiency.
The efficiency $\eta$ enters only in the stochastic
$\mathcal{H}[\cdot]$ term: the monitored fraction of the output
field provides the innovation signal that updates the conditional
state.
In the limit $\eta \to 1$ \emph{and} with intrinsic decoherence
negligible ($\gamma_\mathrm{dec} \ll \eta\gamma$), the stochastic
update fully compensates the measurement-induced dephasing, and
the conditional state can remain nearly pure. For finite
$\grel, \gdeph$, residual decoherence from the unmonitored
intrinsic channels bounds the achievable purity even at $\eta = 1$.
In the limit $\eta \to 0$, no information is extracted, the
$\mathcal{H}[\cdot]$ term vanishes, and the measurement acts as
pure dephasing--recovering the unconditional master equation.

\subsection{Coherence metric and routing decision}

The coherence score $S(t)$ defined in Eq.~\eqref{eq:coherence}
quantifies the $\ell_1$-norm of coherence in the $\sigma_z$ eigenbasis.
It takes values in $[0, 1]$: $S = 0$ for a state diagonal in the
$\sigma_z$ basis (no coherence), and $S = 1$ for a pure equatorial
state such as $\ket{+} = (\ket{0}+\ket{1})/\sqrt{2}$.

At the decision time $T$, the FPGA computes $S(T)$ from the estimated
conditional state and applies the routing rule:
\begin{equation}
    \text{Route} =
    \begin{cases}
        A & \text{if } S(T) > \Sth, \\
        B & \text{otherwise},
    \end{cases}
    \label{eq:routing}
\end{equation}
where $\Sth \in [0,1]$ is a tunable threshold that controls the
tradeoff between certification quality and heralding efficiency.
Higher $\Sth$ demands stronger coherence, accepting fewer trajectories
but certifying higher photon quality.

It is important to emphasize that the conditional coherence $S(t)$ is
not a pre-existing property of the qubit that the measurement merely
reveals. Rather, it is a property of the \emph{Bayesian posterior}:
the best estimate of the qubit state given the measurement record up to
time $t$. The unconditional coherence (obtained by averaging over all
possible measurement records) decays rapidly to a small steady-state
value $S_\mathrm{uncond}^\mathrm{ss} \ll \Sth$ on a timescale
$\sim 1/R_b$, where $R_b = (R_x + R_y)/2$ is the
phase-averaged transverse dephasing rate, with
$R_x = 2\gz + 2\gdeph + \grel/2$ and
$R_y = 2\gx + 2\gz + 2\gdeph + \grel/2$ being the
component-specific rates.
(With the Rabi drive present,
the unconditional steady state retains a small residual coherence from
the driven dynamics; for our parameters,
$S_\mathrm{uncond}^\mathrm{ss} \approx 0.02$.) The conditional coherence, by contrast, can remain of order
unity, because measurement backaction continuously regenerates it on
individual trajectories~\cite{Jacobs2014_Book}. This distinction--between the
unconditional suppression and the trajectory-level persistence of
coherence--is the physical foundation of the protocol, demonstrated
directly in Sec.~\ref{sec:demonstration}.

We also define the conditional purity
\begin{equation}
    \mathcal{P}(t) \equiv \mathrm{Tr}[\rho_c(t)^2]
    = \frac{1 + |\mathbf{r}(t)|^2}{2},
    \label{eq:purity}
\end{equation}
where $|\mathbf{r}|^2 = S^2 + z^2$ is the squared Bloch-vector length,
with $S$ and $z \equiv \langle\sigma_z\rangle_c$ as defined above.
While the coherence score $S$ captures only the equatorial
projection of $\mathbf{r}$, the purity $\mathcal{P}$ depends on the full
length $|\mathbf{r}|$. This distinction underlies a geometric
subtlety in the certification analysis
(Sec.~\ref{sec:geometric_loophole}): an estimator can underestimate
$\mathcal{P}$ while overestimating $S$.

\subsection{Simulation parameters}

All simulations use the parameters in Table~\ref{tab:params}. With
these values, the measurement rates ($\gx/2\pi = \gz/2\pi = 0.1$~MHz)
are ten times larger than the intrinsic decoherence rates
($\grel/2\pi = \gdeph/2\pi = 0.01$~MHz), placing the system in a
regime where measurement-induced dynamics dominate.

\begin{table}[tb]
\centering
\caption{Simulation parameters. Rates are in units of $2\pi$\,MHz;
times are in units of $\mu$s. The decision time $T = 10\;\mu$s and
time step $dt = T/(N_t-1) \approx 5\;$ns.}
\label{tab:params}
\footnotesize
\begin{tabular}{lll}
\toprule
Symbol & Value & Description \\
\midrule
$\Ox / 2\pi$       & 1.0    & Rabi drive strength (MHz) \\
$\Delta / 2\pi$    & 0.5    & Detuning (MHz) \\
$\gx / 2\pi$       & 0.1    & $\sigma_x$ measurement rate (MHz) \\
$\gz / 2\pi$       & 0.1    & $\sigma_z$ measurement rate (MHz) \\
$\grel / 2\pi$     & 0.01   & Relaxation rate (MHz) \\
$\gdeph / 2\pi$    & 0.01   & Pure dephasing rate (MHz) \\
$\eta$             & 0.7    & Measurement efficiency (both channels) \\
$\Sth$             & 0.7    & Routing threshold \\
$T$                & 10.0   & Decision time ($\mu$s) \\
$N_t$              & 2001   & Time steps \\
$N_\mathrm{traj}$  & 3000   & Number of trajectories \\
\bottomrule
\end{tabular}
\end{table}

 The QuTiP
package~\cite{Johansson2012_Qutip,Johansson2013_Qutip2} generates
ground-truth trajectories via \texttt{smesolve}; the same measurement
records are then fed to each estimator for fair comparison.
The pointwise overcertification analysis (Sec.~\ref{sec:pointwise})
uses $N_\mathrm{traj} = 10^6$ Bloch-vector trajectories for improved
statistics.

All trajectories are initialized in a pure equatorial state
$\ket{+} = (\ket{0}+\ket{1})/\sqrt{2}$ (so $S(0) = 1$), and every
estimator is initialized at that same state, so the error process of
Sec.~\ref{sec:pointwise} starts at $E(0) = 0$. Successive routing events use freshly re-initialized
qubits, making rounds conditionally independent and identically
distributed given the device model--an assumption used in the
sequential composition of Sec.~\ref{sec:composable}.

The QuTiP trajectory figures use the angular rates listed in
Table~\ref{tab:params}. The Bloch-vector pointwise analysis of
Sec.~\ref{sec:pointwise} uses the corresponding dimensionless rate
ratios \emph{without} the explicit $2\pi$ factor; this rescales time
and the OU rate parameters $\mu, \sigma_E, \nu$ but leaves the
stationary dimensionless error statistics $\bar{E}$, $\mathrm{Var}(E)$,
and the tail probabilities unchanged.

\section{The measurement--coherence--drive triad}
\label{sec:triad}

\begin{figure*}[t]
    \centering
    \includegraphics[width=\textwidth]{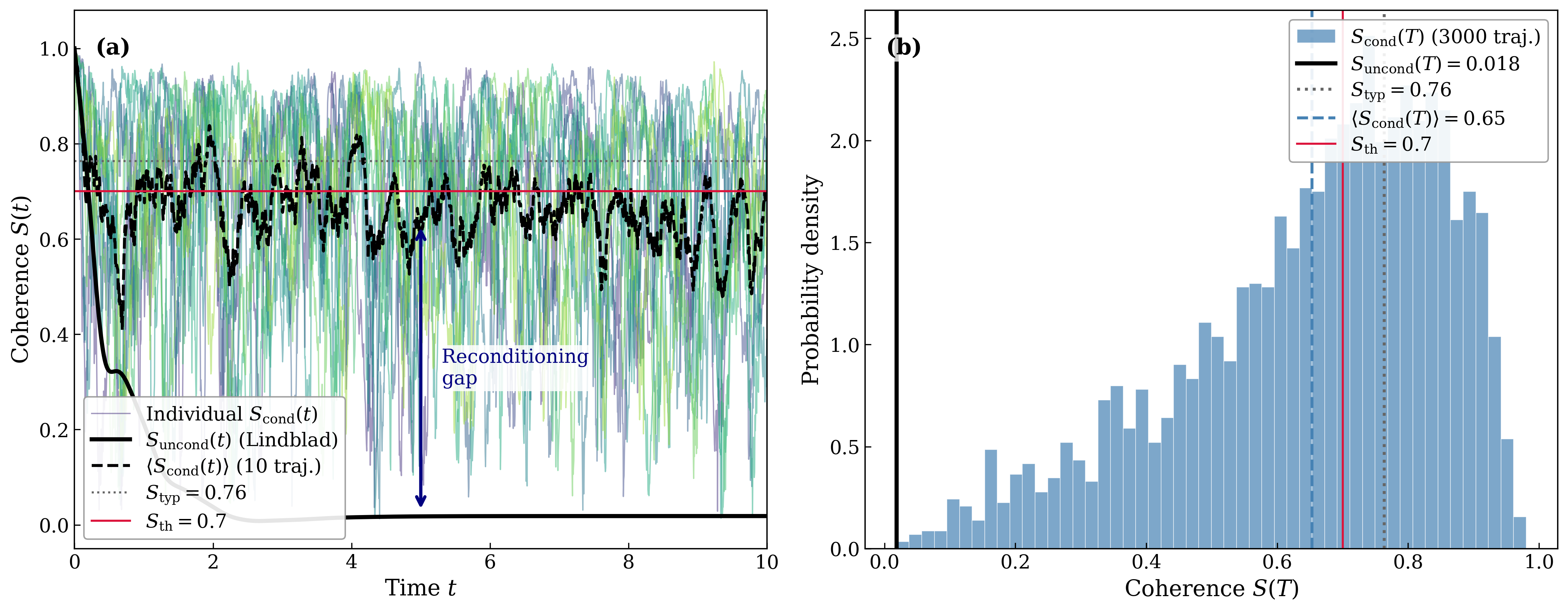}
    \caption{Conditional versus unconditional coherence from $N = 3000$
    trajectories (Table~\ref{tab:params}).
    (a)~Individual conditional trajectories $S_\mathrm{cond}(t)$ (colored)
    oscillate near the analytical estimate $S_\mathrm{typ} = 0.76$
    [dotted; Eq.~\eqref{eq:Styp}]; the dashed black curve is the mean
    over the ten plotted trajectories; the unconditional (Lindblad)
    solution $S_\mathrm{uncond}(t)$ (bold black) decays to
    $\approx 0.02$ within $t \lesssim 2$. The arrow marks the reconditioning gap—the separation between the unconditional decay and the conditional-trajectory envelope.
    (b)~Distribution of $S_\mathrm{cond}(T)$
    over all 3000 trajectories at $t = T$.}
    \label{fig:uncond_vs_cond}
\end{figure*}

The coherence score $S$ defined in Eq.~\eqref{eq:coherence} quantifies the length of the
Bloch vector's projection onto the $x$-$y$ equatorial plane. This
choice of basis is not arbitrary: it is dictated by a self-consistency
requirement between the measurement axes, the coherence basis, and the
Hamiltonian drive, which we now explain.

In our protocol, the two measurement channels ($\sigma_x$ and
$\sigma_z$) introduce dephasing in complementary directions. The
$\sigma_z$ channel dephases the $x$-$y$ components at rate $2\gz$,
while the $\sigma_x$ channel dephases the $y$-$z$ components at rate
$2\gx$. Crucially, both channels are \emph{monitored}: their
backaction is captured by the $\mathcal{H}[\cdot]$ terms in the SME
[Eq.~\eqref{eq:SME}], and the conditional state compensates for the
measurement-induced dephasing. The residual, unrecoverable damage to
coherence comes only from the unmonitored fraction $(1-\eta)$.

Had we instead measured $\sigma_x$ and $\sigma_y$ and defined
coherence in the $\sigma_x$ eigenbasis as
$\tilde{S} = \sqrt{\langle\sigma_y\rangle_c^2 +
\langle\sigma_z\rangle_c^2}$, the $z$-component would appear in the
coherence metric and would suffer dephasing from \emph{both}
measurement channels ($\mathcal{D}[\sigma_x]$ and
$\mathcal{D}[\sigma_y]$ each dephase $z$ at rate $2\gamma$), leading
to larger total measurement backaction on the quantity of interest.

This analysis reveals that three elements of the protocol must be
chosen self-consistently:
\begin{enumerate}
    \item The \textbf{coherence basis} ($\sigma_z$ eigenstates) defines
    which off-diagonal elements of $\rho_c$ constitute the resource to
    be certified.
    \item The \textbf{measurement axes} ($\sigma_x$, $\sigma_z$)
    provide information about the Bloch-vector components while
    introducing dephasing that is complementary to each measurement
    axis.
    \item The \textbf{transverse drive} ($\Ox\sigma_x$)
    continuously regenerates equatorial coherence by driving population
    between $\ket{0}$ and $\ket{1}$, counteracting the
    measurement-induced dephasing of the $x$-$y$ plane.
\end{enumerate}
We refer to this trio as the \emph{measurement--coherence--drive triad}.
Any alternative measurement scheme faces the same
information--disturbance tradeoff: measurement of an observable $O$
provides information about $\langle O \rangle_c$ but dephases the
components complementary to $O$, and coherence is always defined in
terms of those complementary components.
Our particular configuration maps onto the two-tone dispersive scheme
demonstrated by Hacohen-Gourgy \emph{et al.}~\cite{HacohenGourgy2016_Simultaneous},
in which a single readout resonator is driven with two microwave
tones at frequencies chosen so that one tone's dispersive shift
couples to $\sigma_z$ while the stroboscopic average of the other
couples to $\sigma_x$ in the rotating frame. The two channels are
separated by demodulation of the reflected signal at the respective
sideband frequencies. Transverse microwave drives implement the
Rabi term $\Ox\sigma_x$ via standard single-qubit control. This
realization is not merely conceptual: simultaneous $\sigma_x$--$\sigma_z$
monitoring has been demonstrated at efficiencies $\eta \sim
0.4$--$0.5$ in current hardware~\cite{HacohenGourgy2016_Simultaneous},
though our protocol's quantitative guarantees (Table~\ref{tab:composable})
assume $\eta = 0.7$--achievable with improved parametric amplifier
chains and reduced circulator losses.

\subsection{Conditional versus unconditional coherence: Simulation}
\label{sec:demonstration}

Before developing applications, we establish empirically the central
claim of the protocol: the conditional coherence $S_\mathrm{cond}(t)$
on individual quantum trajectories remains large and volatile, hovering
near the analytical estimate $S_\mathrm{typ}$, while the unconditional
(Lindblad) coherence $S_\mathrm{uncond}(t)$ collapses irreversibly to
near zero. This gap between the trajectory-level and ensemble-level
pictures is the physical foundation of coherence-gated routing.

The reference scale $S_\mathrm{typ}$ follows from a two-step balance
argument. Equating the purification rate of the monitored channels
(which scales as $\eta\gamma$; cf.\ Proposition~1) against the
intrinsic decoherence rate $\gamma_\mathrm{dec}$ gives a stationary
Bloch length $|\mathbf{r}|_\mathrm{ss}^2 \approx
\eta\gamma/(\eta\gamma + \gamma_\mathrm{dec})$; treating the Bloch
direction as approximately isotropic under the combined action of the
drive and the dual-axis backaction then gives
$\langle S^2\rangle = \tfrac{2}{3}\langle|\mathbf{r}|^2\rangle$, since
$\langle\sin^2\theta\rangle_\mathrm{iso} = 2/3$. Together,
\begin{equation}
    S_\mathrm{typ} \approx
    \sqrt{\frac{2\,\eta\gamma}{3\,(\eta\gamma + \gamma_\mathrm{dec})}}
    \approx 0.76
    \label{eq:Styp}
\end{equation}
for the parameters of Table~\ref{tab:params}, with
$\gamma = \gx + \gz$ the total measurement rate. Both steps are
heuristic--the direction distribution is not exactly isotropic, and
the purity balance neglects the drive--so $S_\mathrm{typ}$ is a
typical scale, not a predicted mean. Its $|\mathbf{r}| \to 1$ limit,
$\sqrt{2/3} \approx 0.82$, is the isotropic ceiling quoted in
Sec.~\ref{sec:discussion}.

Figure~\ref{fig:uncond_vs_cond}(a) shows ten representative conditional
trajectories alongside the Lindblad solution. The individual
$S_\mathrm{cond}(t)$ traces remain of order unity throughout the
evolution, regularly crossing the threshold $\Sth = 0.7$ and enabling
certification. The Lindblad solution decays to a near-zero steady state
$S_\mathrm{uncond}^\mathrm{ss} \approx 0.02$.

Figure~\ref{fig:uncond_vs_cond}(b) shows the final-time distribution
$P[S_\mathrm{cond}(T)]$ across 3000 trajectories. The distribution is
broad and weighted toward high coherence, with the empirical mean
$\langle S_\mathrm{cond}(T)\rangle \approx 0.65$ sitting somewhat
below the analytical estimate $S_\mathrm{typ} = 0.76$, and a
substantial fraction of events falling to the right of $\Sth = 0.7$.
The routing decision therefore selects from a genuine population of
high-coherence events, not from a uniformly degraded ensemble.
The reconditioning mechanism behind this gap is unpacked further in
Supplementary Information Sec.~S2.
\newline

\section{Certified quantum random number generation}
\label{sec:qrng}

We now turn to the first application of coherence-gated routing:
certified quantum random number generation. The key observation is
that the Bloch-sphere geometry imposes a rigid constraint between the
coherence score $S$ and the $z$-component of the conditional Bloch
vector, $z(T) \equiv \langle\sigma_z\rangle_c(T)$, which governs the
Born-rule outcome statistics.

\subsection{Min-entropy bound from Bloch-sphere geometry}

For a trajectory accepted at threshold $\Sth$, the geometric constraint
$|z| \leq z_{\max} \equiv \sqrt{1 - \Sth^2}$ bounds the Born-rule
probability:
\begin{equation}
    P_z \in \left[\frac{1 - z_{\max}}{2},\; \frac{1 + z_{\max}}{2}\right].
    \label{eq:Pz_bound}
\end{equation}
The worst-case (minimum) min-entropy per bit is therefore:
\begin{equation}
    \boxed{H_\infty(\Sth) = -\log_2\!\left(\frac{1 + \sqrt{1 - \Sth^2}}{2}
    \right).}
    \label{eq:Hmin}
\end{equation}
At $\Sth = 0.95$, this gives $H_\infty \geq 0.61$ bits per event;
at $\Sth = 0.99$, $H_\infty \geq 0.81$ bits. Intermediate
thresholds are tabulated in Table~S2 (Supplementary Information
Sec.~S9).

The certified random bit itself is produced by a terminal projective
measurement of $\sigma_z$ on the accepted qubit at the decision time
$T$, returning outcome $\pm 1$ with Born-rule probability
$P_z = (1 \pm z(T))/2$; Eq.~\eqref{eq:Hmin} lower-bounds the
min-entropy of precisely this outcome. The continuous dual-axis
monitoring plays a purely certifying role: the measurement record up to
$T$ establishes $S(T) \geq \Sth$, which through the geometric
constraint $|z(T)| \leq \sqrt{1-\Sth^2}$ makes Eq.~\eqref{eq:Pz_bound}
hold. Certification and bit generation therefore occur within the same
measurement cycle on the same qubit, with no separate test rounds.

\subsection{Numerical validation}

\begin{figure*}[t]
    \centering
    \includegraphics[width=\textwidth]{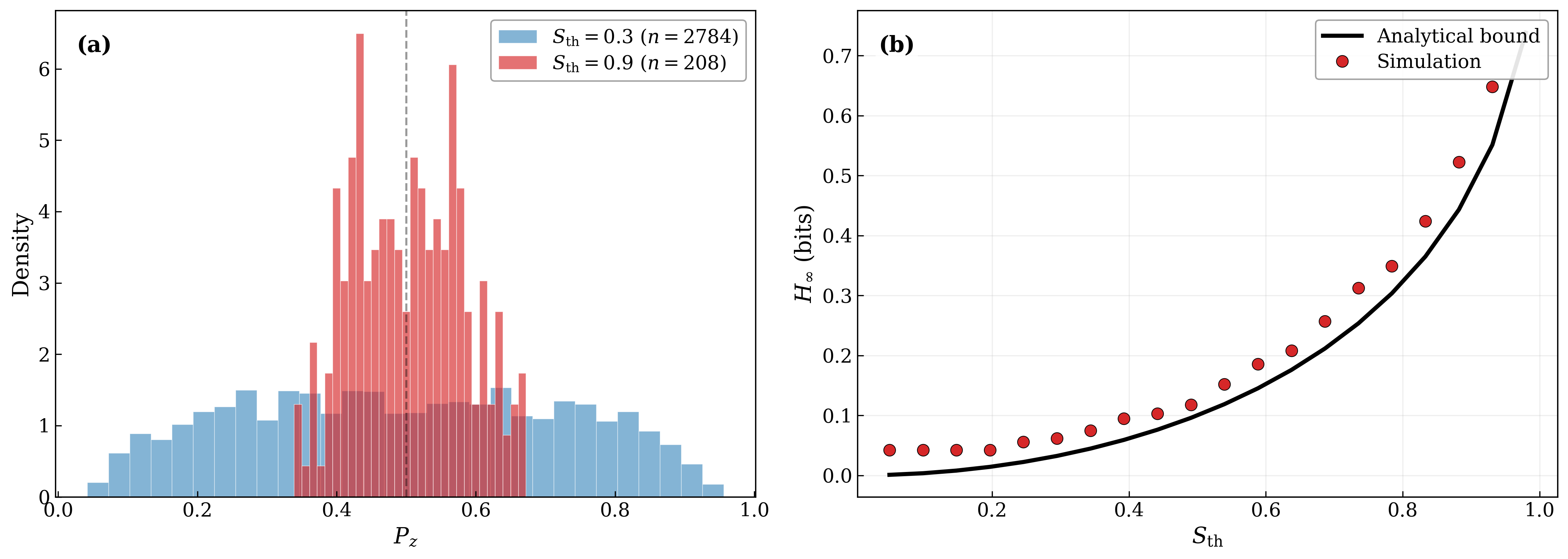}
    \caption{QRNG certification from $N = 3000$ trajectories.
    (a)~Distribution of $P_z$ at
    $\Sth = 0.3$ (broad) and $\Sth = 0.9$ (narrow).
    (b)~Min-entropy: empirical values (markers) lie above
    the analytical bound (solid) at all thresholds.
    Empirical points at high $\Sth$ rest on few accepted events and
    carry correspondingly large statistical uncertainty (see text).}
    \label{fig:qrng}
\end{figure*}

\begin{figure}[tb]
    \centering
    \includegraphics[width=\columnwidth]{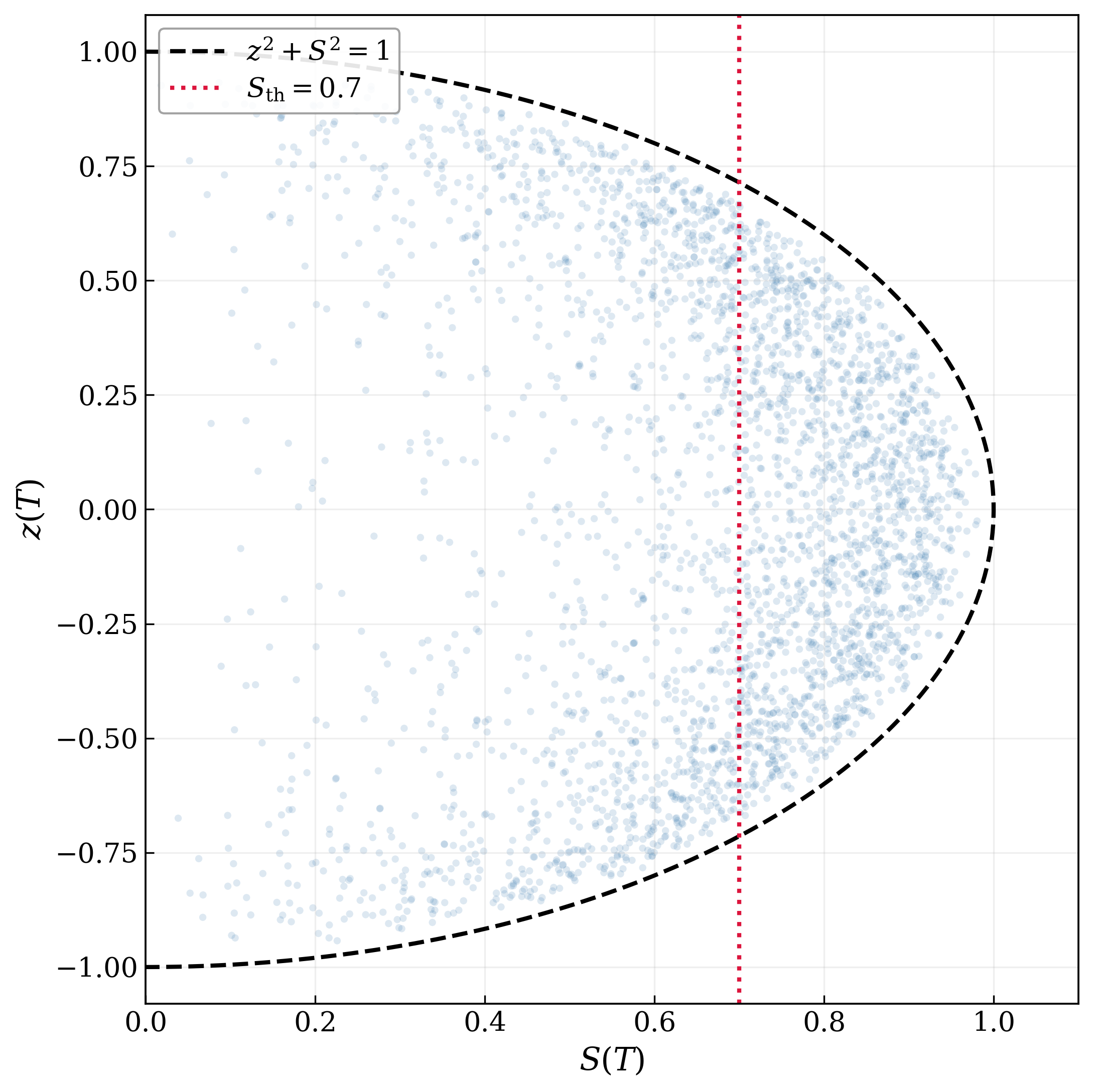}
    \caption{Bloch-sphere diagnostic: $S(T)$ vs.\
    $z(T) = \langle\sigma_z\rangle_c(T)$ across 3000 trajectories.
    All points lie within the Bloch-sphere boundary $S^2 + z^2 \leq 1$
    (dashed curve).}
    \label{fig:bloch_diagnostic}
\end{figure}

Figures~\ref{fig:qrng} and~\ref{fig:bloch_diagnostic} validate the
certification mechanism. In Fig.~\ref{fig:qrng}(a), the distribution
of $P_z$ narrows sharply around $0.5$ as $\Sth$ increases.

Panel~(b) confirms that the empirical
min-entropy lies above the analytical bound at all thresholds.
At high thresholds the empirical estimate rests on few accepted events
($n = 208$ at $\Sth = 0.9$, fewer still at $\Sth = 0.95$), so the
markers carry substantial finite-sample uncertainty; we therefore read
panel~(b) as a consistency check on the geometric bound rather than as
an independent certification of it.
A per-trajectory view of the heralding decision--coherence and
Born-rule probability shown separately for accepted and rejected
events--appears in Fig.~S2 (Supplementary Information Sec.~S4). 
Figure~\ref{fig:bloch_diagnostic} verifies that all simulated points 
lie inside $S^2+z^2 \le1$, so the geometric min-entropy bound is applied 
within the physical Bloch region.

\subsection{Security role of the estimator}

The min-entropy guarantee of Eq.~\eqref{eq:Hmin} holds only if the
coherence estimate is faithful--specifically, only if accepting at
threshold $\Sth$ genuinely certifies
$S_\mathrm{true}(T) \geq \Sth - \varepsilon$ up to a controlled
failure probability. This requirement is analyzed in
Secs.~\ref{sec:certification}--\ref{sec:pointwise}, where we develop
both an average-case argument (Proposition~1) and pointwise tail
bounds (Theorem~2 and the OU comparison).

Three scoping remarks sharpen the claim. First, the geometric bound
Eq.~\eqref{eq:Pz_bound} constrains the Born-rule probability of the
conditional state given the full measurement record, so the resulting
min-entropy is conditional on everything the FPGA knows--in
particular, publishing the classical heralding message
$(\varphi, S(T))$ leaks nothing further about the bit. Second, the
guarantee is against adversaries described by the device model: an
adversary holding a purification of the unmonitored environment
fraction $(1-\eta)$ lies outside the SME description, and excluding
such adversaries is precisely the model-certification assumption~(i)
of Sec.~\ref{sec:qrng_comparison}. Third, rounds compose sequentially:
each event uses a freshly initialized qubit (Sec.~\ref{sec:protocol}),
so $n$ accepted events certify $n H_\infty$ bits except with
probability at most $n\,\delta(\Sth, \varepsilon)$ by the union bound, with
$\delta(\Sth, \varepsilon)$ the per-event failure probability of
Sec.~\ref{sec:composable}.

\subsection{Comparison with existing QRNG protocols}
\label{sec:qrng_comparison}

Quantum random number generators span a hierarchy of trust
assumptions~\cite{Herrero2017_QRNG_Review,Ma2016_QRNG}. At one extreme,
\emph{fully device-independent} (DI) QRNGs derive randomness from
Bell-inequality violation~\cite{Pironio2010_DI_QRNG,Liu2018_DI_QRNG}.
At the other,
\emph{trusted-device} QRNGs assume full knowledge of the
source~\cite{Jennewein2000_QRNG,Symul2011_QRNG}. Our coherence-certified
QRNG occupies intermediate ground.

The min-entropy bound [Eq.~\eqref{eq:Hmin}] requires three conditions:
(i)~the SME [Eq.~\eqref{eq:SME}] is a correct model of the
qubit--measurement dynamics;
(ii)~the measurement efficiency $\eta$ is calibrated to within the
conservative margin provided by $\etaa < \etatrue$;
(iii)~the estimator's coherence estimate does not
systematically overcertify the true conditional coherence.
With the composable bound developed in Sec.~\ref{sec:pointwise},
condition~(iii) is quantified through the joint acceptance--failure
event:
$\Pr[\hat{S}(T) > \Sth \,\wedge\, S_\mathrm{true}(T) < \Sth -
\varepsilon] \leq \delta(\Sth, \varepsilon)$, where $\delta(\Sth, \varepsilon)$ is
the failure probability defined in Eq.~\eqref{eq:delta_def} and
tabulated for representative operating points in
Table~\ref{tab:composable}; the OU bound of Eq.~\eqref{eq:OU_bound} is
the special case $\varepsilon = 0$.
This places our protocol in a
\emph{model-certified} regime--stronger than a fully trusted device
but weaker than semi-device-independent protocols that certify randomness
from dimension witnesses alone~\cite{Li2011_SemiDI,Bowles2014_DimensionWitness,Lunghi2015_SemiDI}.

Two features are specific to our approach: (i)~certification and
bit generation occur in the same measurement cycle--no separate test
rounds are needed; (ii)~the conservative-$\eta$ estimator
($\etaa < \etatrue$) suppresses overcertification, as quantified in
Secs.~\ref{sec:certification}--\ref{sec:pointwise}.

\section{Coherence-certified photon sources for quantum networks}
\label{sec:qnet}

The second application serves as a
certified photon source for quantum networks.
When a photon is accepted
at high $\Sth$, the network node receives not only the photon but also
a classical side-channel message from the FPGA containing the
equatorial angle
$\varphi = \arctan(\langle\sigma_y\rangle_c /
\langle\sigma_x\rangle_c)$ and the coherence $S(T)$.

\subsection{Phase-aware feedforward}

\begin{figure}[tb]
    \centering
    \includegraphics[width=\columnwidth]{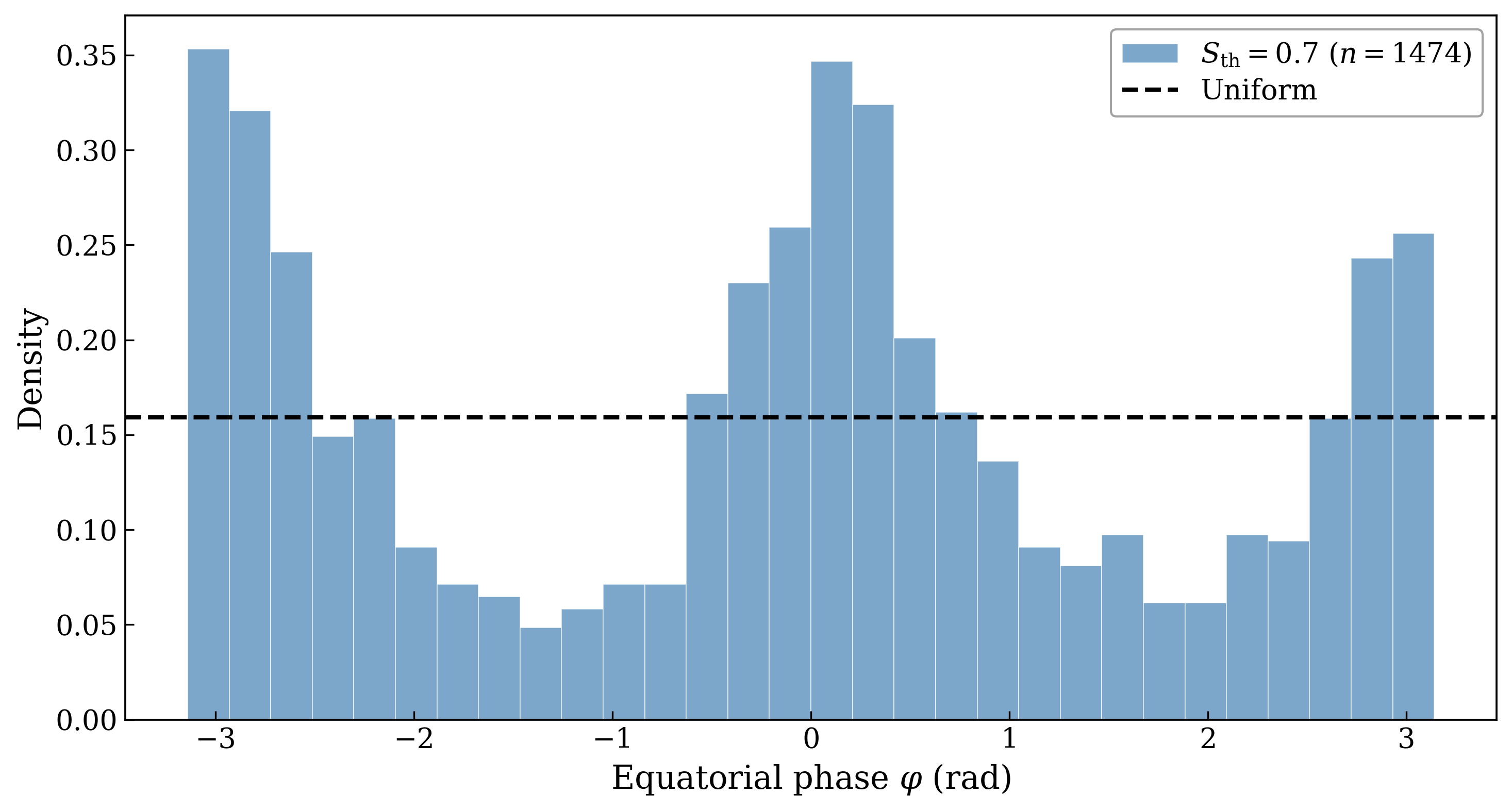}
    \caption{Distribution of the equatorial phase angle
    $\varphi$ among $n = 1474$ trajectories accepted
    at $\Sth = 0.7$. Feedforward correction
    $R_z(-\varphi)$ removes this per-trajectory phase. The dashed line marks a uniform phase
distribution.}
    \label{fig:phase}
\end{figure}

The equatorial angle $\varphi$ differs for each trajectory because the
interplay between the deterministic Rabi drive and the stochastic
measurement backaction produces a different phase evolution on each run
(Fig.~\ref{fig:phase}). This is a crucial point of departure from
conventional heralded sources, where phase information about the
emitted photon is either absent or available only as an ensemble
average. Without correction, the per-trajectory phase scatter would
appear as a random global phase on the photon--harmless for some
applications but fatal for any protocol that relies on photons from
different runs being mutually coherent, such as Hong--Ou--Mandel
interference at a Bell-state measurement station. Since the FPGA
already knows $\varphi$ from its real-time state estimate, it applies
a compensating $R_z(-\varphi)$ rotation before the photon enters the
network. The cost is one digital phase update per accepted event;
the benefit is that successive accepted photons are aligned in the
equatorial plane to a common reference frame. This phase alignment
is what makes the entanglement-distribution analysis in
Sec.~\ref{sec:entanglement} possible at all: independent nodes can
produce photons whose Bloch vectors point along the same axis without
ever exchanging quantum information, using only the classical
side-channel that already accompanies the heralding signal.

\subsection{Quality-of-service routing}

The tunable threshold $\Sth$ allows a single physical device to serve
multiple protocols at different points on the
quality--rate tradeoff curve, simply by adjusting $\Sth$ in software.
Consider a deployment where the same hardware feeds three downstream
clients: a QRNG that needs maximally certified bits but tolerates a
low rate; a quantum-key-distribution channel that needs moderate
fidelity at moderate rate; and a clock-synchronization or bit-commitment
service that needs only mild coherence certification but high
throughput. Conventional heralded sources are typically optimized at
fabrication for a single operating point. A coherence-gated router
moves these knobs into the digital control loop. Setting $\Sth = 0.95$
gives ${\sim}2\%$ heralding efficiency with ${\geq}0.61$ bits of
certified min-entropy per event; lowering it to $\Sth = 0.7$ trades
that down to ${\sim}50\%$ heralding at ${\geq}0.22$ bits (both quoting
the $\varepsilon = 0$ geometric values of Eq.~\eqref{eq:Hmin};
finite-$\varepsilon$ composable values appear in
Table~\ref{tab:composable}). The change is
a one-line FPGA register write. This kind of dynamic
reconfiguration matters for systems integration in a quantum network,
where different time slots may serve different protocols and the
operating point should track demand rather than be locked at
build time.

\subsection{Multiplexed operation}

Operating at a high coherence threshold reduces the per-module
heralding rate, but multiplexing recovers throughput. With $N$
independent coherence-gated routers running in parallel and a
heralding efficiency $\eta_h$ at threshold $\Sth$, the aggregate rate
of certified photons is
\begin{equation}
    R_\mathrm{eff} = \frac{N \eta_h}{T},
    \label{eq:multiplexed_rate}
\end{equation}
where $T$ is the decision time per event. For $N = 10$ modules at
$\eta_h \approx 7\%$ ($\Sth \approx 0.9$) and $T = 10\;\mu$s, this
gives $R_\mathrm{eff} \sim 7\times10^4$ certified photons per
second--comparable with current heralded single-photon
sources~\cite{MeyerScott2020}. The architectural advantage is that
each module is a self-contained transmon qubit with its own
two-tone monitoring chain and FPGA, so adding modules scales the
rate linearly without requiring any quantum communication between
them. The classical side-channel from each module's FPGA can be
multiplexed onto a shared digital backplane, leaving the quantum
channels independent. This separation between the per-module
quantum complexity and the system-level classical aggregation is
what makes the platform attractive at scale: scaling out is a
classical engineering problem, not a quantum-coherence one.

\section{Coherence-bounded entanglement distribution}
\label{sec:entanglement}

\begin{figure}[tb]
\centering
\begin{tikzpicture}[x=0.72cm, y=0.72cm, >=Latex,
    box/.style={draw, thick, rounded corners=2pt, minimum height=0.6cm,
                inner sep=3pt, font=\footnotesize},
    circ/.style={draw, thick, circle, minimum size=0.5cm,
                 font=\footnotesize}]
\node[box, fill=blue!10, minimum width=1.8cm] (NA) at (-4,0) {Node A};
\node[font=\scriptsize] at (-4,-0.7) {$S_A \geq \Sth$};
\node[box, fill=red!10, minimum width=1.8cm] (NB) at (4,0) {Node B};
\node[font=\scriptsize] at (4,-0.7) {$S_B \geq \Sth$};
\node[box, fill=green!10, minimum width=1.5cm] (BSM) at (0,0) {BSM};
\draw[->, thick, blue!60] (NA) -- node[above, font=\tiny] {$\gamma_A$} (BSM);
\draw[->, thick, red!60] (NB) -- node[above, font=\tiny] {$\gamma_B$} (BSM);
\draw[->, dashed] (NA) -- ++(0,-1.5) -| node[below, font=\tiny, pos=0.25]
     {$S_A, \varphi_A$} (BSM);
\draw[->, dashed] (NB) -- ++(0,-1.5) -| node[below, font=\tiny, pos=0.25]
     {$S_B, \varphi_B$} (BSM);
\end{tikzpicture}
\caption{Two-node entanglement distribution protocol. Each node
performs a controlled qubit--cavity interaction that produces a
qubit--photon Bell pair; the photon is sent to the central BSM
station while the matter qubit is retained. A successful BSM
projects the two remote matter qubits into an entangled state.}
\label{fig:entanglement}
\end{figure}
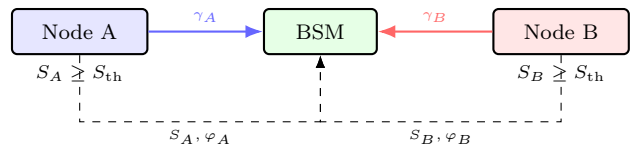

Consider two remote nodes (Fig.~\ref{fig:entanglement}), each running
the coherence-gated router independently under four idealizing
assumptions (A1--A4; see Supplementary Information Sec.~S1 for
details).
We emphasize that this section is a forward-looking application
sketch rather than a demonstrated networking protocol: the analysis
isolates how coherence certification constrains the achievable
\emph{matter--matter entanglement fidelity} after a downstream
Bell-state measurement (BSM), but does not model the full photonic
transduction stack (finite gate fidelity, BSM efficiency $\leq 1/2$,
photon indistinguishability) that would determine the final
entanglement rate and fidelity in practice. Relaxing A1--A4
introduces additional loss factors that degrade the bounds; the
product structure of Eq.~\eqref{eq:Fent} is preserved for
independent multiplicative imperfections (e.g., finite $\Fgate$),
though correlated errors such as photon distinguishability in the
BSM may modify the functional form.

After feedforward correction $R_z(-\varphi_i)$, node~$i$ performs a
controlled qubit--cavity interaction (e.g., conditional photon
emission~\cite{Kurpiers2018} or a transmon--resonator
iSWAP~\cite{Zhong2021}) that maps the data qubit's state into a
qubit--photon Bell pair:
\begin{equation}
    \alpha\ket{0}_q + \beta\ket{1}_q \;\longrightarrow\;
    \alpha\ket{0}_q\ket{0}_\gamma + \beta\ket{1}_q\ket{1}_\gamma.
    \label{eq:bell_pair}
\end{equation}
Crucially, the matter qubit \emph{retains} its half of the
entanglement; the photon $\gamma$ is emitted into the transmission
line.

To see how coherence certification controls the Bell-pair quality,
consider the idealized case: after phase correction, the conditional
state at the moment of emission is
$\rho_i = \frac{1}{2}\bigl(\mathbb{I} + S_i\,\sigma_x
+ z_i\,\sigma_z\bigr)$.
For a trajectory accepted at threshold $\Sth$, the coherence
$S_i \geq \Sth$ dominates the Bloch vector (since
$|z_i| \leq \sqrt{1-S_i^2}$ is geometrically suppressed).
The fidelity of the resulting Bell pair
[Eq.~\eqref{eq:bell_pair}] with the ideal maximally entangled state
$\ket{\Phi^+} = (\ket{00}+\ket{11})/\sqrt{2}$ is
\begin{equation}
    F_i = \bra{\Phi^+}
    \bigl(\rho_i \otimes \ket{0}\!\bra{0}\bigr)_{\!\mathrm{ent.\,map}}
    \ket{\Phi^+}
    = \frac{1 + S_i}{2}
    \geq \frac{1 + \Sth}{2},
    \label{eq:Fi}
\end{equation}
where the middle equality is seen in one line: the entangling map of
Eq.~\eqref{eq:bell_pair} is the isometry
$U: \ket{0}_q\ket{0}_\gamma \mapsto \ket{00}$,
$\ket{1}_q\ket{0}_\gamma \mapsto \ket{11}$, for which
$U^\dagger\ket{\Phi^+} = \ket{+}_q\ket{0}_\gamma$, so that
$F_i = \bra{+}\rho_i\ket{+}
= \tfrac{1}{2}\bigl(1 + \langle\sigma_x\rangle_{\rho_i}\bigr)$.
After the feedforward rotation the equatorial component of $\rho_i$
points along $+x$, so $\langle\sigma_x\rangle = S_i$ and
$\langle\sigma_y\rangle = 0$. The longitudinal component $z_i$ drops
out exactly, because $\bra{+}\sigma_z\ket{+} = 0$: no explicit $z_i$
penalty appears in $F_i$ even though
$|z_i| \leq \sqrt{1 - S_i^2}$ is generally nonzero.

\begin{figure*}[t]
    \centering
    \includegraphics[width=\textwidth]{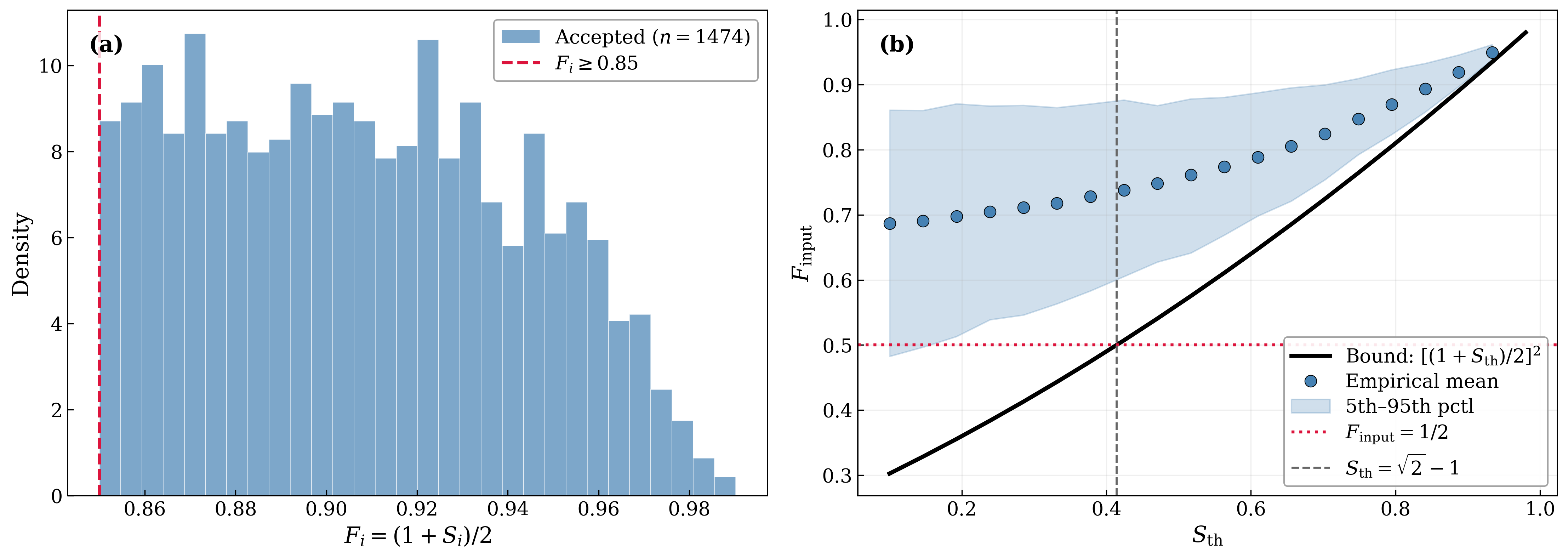}
    \caption{Input product-state quality (ideal $\Fgate = 1$).
    (a)~Single-node Bell-pair fidelity distribution at $\Sth = 0.7$.
    (b)~$F_\mathrm{input}$ vs.\ $\Sth$; empirical (markers)
    above the analytical bound (solid).}
    \label{fig:entanglement_fidelity}
\end{figure*}

For an ideal entangling gate ($\Fgate = 1$), the
input product-state fidelity available to the downstream BSM is
\begin{equation}
    F_\mathrm{input} = F_A \cdot F_B
    \geq \left(\frac{1 + \Sth}{2}\right)^2.
    \label{eq:Fent}
\end{equation}
A successful BSM on the two photons projects the retained matter
qubits at nodes~A and~B into an entangled state, whose fidelity with
the target Bell state we denote $\Fmm$. It is tempting to suppose that
the swap can only improve on the input product fidelity, i.e.\
$\Fmm \geq F_\mathrm{input}$; this is \emph{not} correct, and we
establish the correct floor here. For the post-feedforward conditional
states $\rho_i = \tfrac12(\mathbb{I} + S_i\,\sigma_x + z_i\,\sigma_z)$,
an ideal BSM (assumption~A3) heralding the $\ket{\Phi^+}$ photon
outcome gives the exact matter--matter fidelity (derived in
Supplementary Information Sec.~S1)
\[
    \Fmm = \frac12 + \frac{S_A S_B}{2\,(1 + z_A z_B)}.
\]
The longitudinal components $z_i$ \emph{survive the swap}, even though
they drop out of the single-node fidelity $F_i$ (because
$\bra{+}\sigma_z\ket{+} = 0$) and are untouched by the feedforward
$R_z(-\varphi_i)$, which fixes only the equatorial angle. The BSM
correlates the two nodes' $z$-components, so for aligned tilts
($z_A z_B > 0$) the matter--matter fidelity falls \emph{below} the
input product $F_\mathrm{input} = F_A F_B$ of Eq.~\eqref{eq:Fent}.
Because certification constrains only $S_i \geq \Sth$, leaving
$|z_i| \leq \sqrt{1-\Sth^2}$ otherwise free, the guaranteed floor is the
worst case over that region, attained at $S_i = \Sth$ with aligned
$z_i = \sqrt{1-\Sth^2}$:
\[
    \boxed{\;\Fmm \;\geq\; \frac12 + \frac{\Sth^2}{2\,(2-\Sth^2)}\;.}
\]
This floor replaces the previously asserted $[(1+\Sth)/2]^2$, which is
a valid lower bound only for $\Sth \lesssim 0.53$. At $\Sth = 0.7$, for
instance, two certified nodes with $S_i = 0.7$ and aligned
$z_i = \sqrt{0.51}$ give $\Fmm = 0.662$, well below the
$[(1+0.7)/2]^2 = 0.7225$ that the product form would predict. For
imperfect entangling gates with $\Fgate < 1$ the floor acquires a
leading-order multiplicative prefactor $\Fgate^2$ (current circuit-QED
implementations achieve $\Fgate \approx
0.97$~\cite{Kurpiers2018,Zhong2021}), which factorizes cleanly from the
coherence-certification contribution. The single-node, input-product,
and matter--matter floors at representative thresholds are collected in
Table~\ref{tab:entanglement}. Figure~\ref{fig:entanglement_fidelity}
shows the input product-state fidelity $F_\mathrm{input}$
[Eq.~\eqref{eq:Fent}] at $\Sth = 0.7$ together with the threshold
sweep; we stress that this figure validates the \emph{pre}-BSM input
fidelity against its bound, not the post-swap floor above.

\begin{table*}[t]
\centering
\caption{Fidelity bounds under idealized BSM ($\Fgate = 1$; see text
for the finite-$\Fgate$ correction). $F_i \geq (1+\Sth)/2$ is the
single-node Bell-pair fidelity [Eq.~\eqref{eq:Fi}] and
$F_\mathrm{input} \geq [(1+\Sth)/2]^2$ [Eq.~\eqref{eq:Fent}] the
\emph{pre}-BSM input product, while
$\Fmm \geq \tfrac12 + \Sth^2/[2(2-\Sth^2)]$ is the worst-case
\emph{post}-BSM matter--matter floor (the $\varepsilon = 0$ value of the
boxed bound in Sec.~\ref{sec:entanglement}), attained at aligned
threshold-coherence pure states. $R_\mathrm{2\text{-}node}$ is the ideal
aggregate two-node coincidence rate for $N = 10$ paired multiplexed
modules per node [scaling $R \sim N \eta_h^2/T$,
Eq.~\eqref{eq:multiplexed_rate}, with $T = 10\;\mu$s], before BSM
efficiency ($\leq 1/2$ for linear optics) and channel losses. Slotted,
synchronized operation is assumed: a coincidence requires both modules
of a pair to herald in the same $T$-slot.}
\label{tab:entanglement}
\begin{tabular}{cccccc}
\toprule
$\Sth$ & $F_i \geq$ & $F_\mathrm{input} \geq$ & $\Fmm \geq$ & $\eta_h$ &
$R_\mathrm{2\text{-}node}$ (ideal) \\
\midrule
0.50 & 0.750 & 0.563 & 0.571 & $\sim 80\%$ & $\sim 6 \times 10^5$/s \\
0.70 & 0.850 & 0.722 & 0.662 & $\sim 50\%$ & $\sim 3 \times 10^5$/s \\
0.90 & 0.950 & 0.903 & 0.840 & $\sim 7\%$ & $\sim 5 \times 10^3$/s \\
0.95 & 0.975 & 0.951 & 0.911 & $\sim 2\%$ & $\sim 4 \times 10^2$/s \\
\bottomrule
\end{tabular}
\end{table*}

With the composable security framework of Sec.~\ref{sec:pointwise},
these bounds acquire explicit failure probabilities: except with
probability at most $1 - (1-\delta(\Sth, \varepsilon))^2 \leq
2\,\delta(\Sth, \varepsilon)$, a coincidence of accepted heralds certifies
$\Fmm \geq \tfrac12 + (\Sth-\varepsilon)^2/[2\,(2-(\Sth-\varepsilon)^2)]$ (Corollary~2).

\section{Certification integrity: the estimator as security primitive}
\label{sec:certification}

The applications developed above depend critically on the faithfulness
of the coherence estimate. An estimator that systematically
overestimates $S(T)$ compromises the QRNG min-entropy bound,
degrades Bell-pair fidelity for networking, and weakens the
matter--matter entanglement fidelity $\Fmm$ for entanglement
distribution.
Conversely, an estimator
that underestimates $S(T)$ is conservative: it rejects good
trajectories but rarely falsely certifies--the residual rate is
itself bounded by the pointwise failure probability $\delta(\Sth, \varepsilon)$
of Sec.~\ref{sec:pointwise}.

\begin{figure*}[t]
    \centering
    \includegraphics[width=\textwidth]{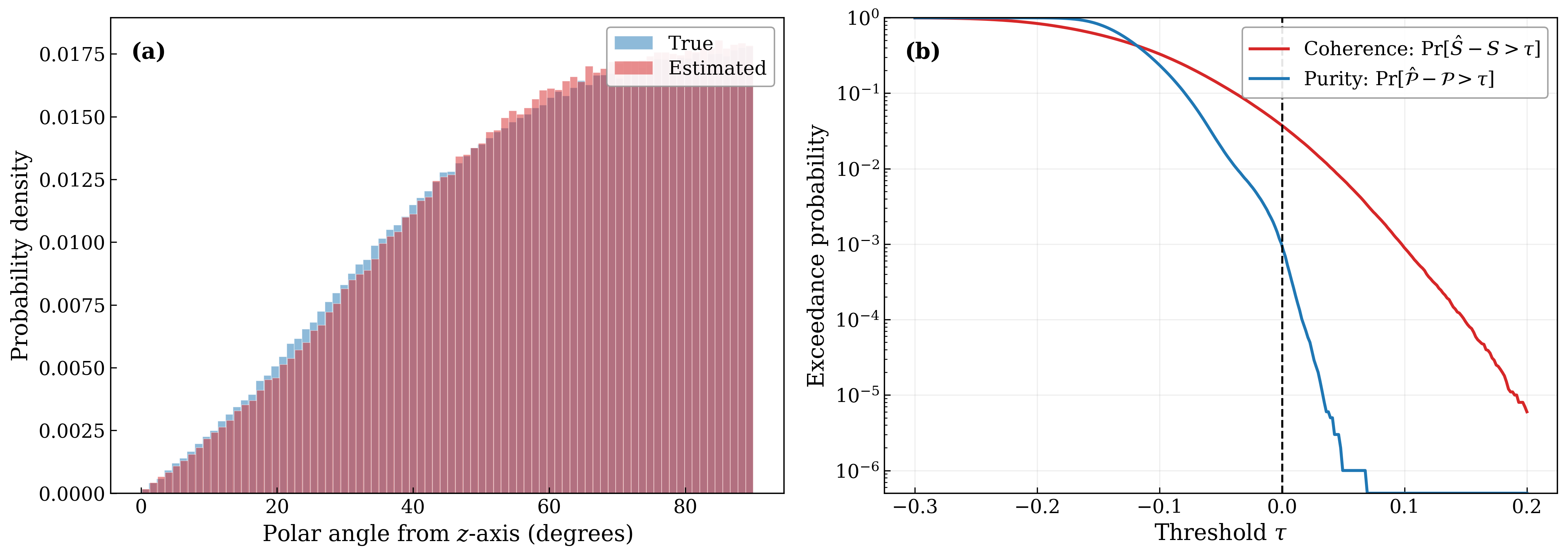}
    \caption{Geometric mechanism of overcertification from $10^6$
    trajectories.
    (a)~Polar angle distribution: the estimated
    state (red) is slightly more equatorial than the true state (blue).
    (b)~Exceedance probability for purity (blue) and coherence (red)
    differences. At threshold $\tau = 0$, coherence
    overcertification ($3.7\%$) exceeds purity overcertification
    ($0.1\%$) by more than an order of magnitude (${\sim}40\times$).}
    \label{fig:geometric}
\end{figure*}

\subsection{Purity-production monotonicity}

\noindent\textbf{Proposition 1} (Monotone purity production).
\textit{For a qubit monitored via homodyne channels $k \in \{x,z\}$
with strengths $\gamma_k$ and common efficiency $\eta$, the explicit
$\eta$-dependence of the ensemble-averaged purity production rate is
strictly positive at every state:}
\begin{equation}
    \left.\frac{\partial}{\partial\eta}\,
    \frac{d\,\mathbb{E}[\mathcal{P}]}{dt}\right|_{\mathrm{expl}}
    = 2\sum_{k=x,z} \gamma_k\,
      \mathbb{E}\!\left[\,\bigl|\mathbf{e}_k - r_k\,\mathbf{r}\bigr|^2
      \right] > 0,
    \label{eq:purity_mono}
\end{equation}
\textit{where $\mathbf{e}_k$ is the unit vector along axis $k$ and
$|\mathbf{e}_k - r_k\mathbf{r}|^2
= (1-r_k^2)^2 + r_k^2\,(|\mathbf{r}|^2 - r_k^2)$.}

\medskip
\noindent\textit{Proof.}
For a qubit, $\mathcal{H}[\sigma_k]\rho =
(\mathbf{e}_k - r_k\mathbf{r})\cdot\boldsymbol{\sigma}$, as follows
from $\{\sigma_k, \rho\} = \sigma_k + r_k\mathbb{I}$. Applying
It\=o's formula to $\mathcal{P} = \mathrm{Tr}(\rho_c^2)$ under the
SME, the quadratic-variation term of channel $k$ contributes
$\eta\gamma_k\,
\mathrm{Tr}\!\bigl[(\mathcal{H}[\sigma_k]\rho_c)^2\bigr]\,dt
= 2\eta\gamma_k\,|\mathbf{e}_k - r_k\mathbf{r}|^2\,dt$
(using $\mathrm{Tr}[(\mathbf{a}\cdot\boldsymbol{\sigma})^2] =
2|\mathbf{a}|^2$); the innovation term is a martingale increment with
zero mean; and all dissipator contributions are independent of
$\eta$. Differentiating with respect to $\eta$ gives
Eq.~\eqref{eq:purity_mono}. Strict positivity:
$|\mathbf{e}_x - r_x\mathbf{r}|^2 = 0$ forces
$\mathbf{r} = \pm\mathbf{e}_x$, at which
$|\mathbf{e}_z - r_z\mathbf{r}|^2 = 1$; no qubit state is a
simultaneous eigenstate of $\sigma_x$ and $\sigma_z$.
\hfill$\square$

\medskip
\noindent
For pure states ($|\mathbf{r}| = 1$) the integrand reduces to
$\mathrm{Var}_c(\sigma_k) = 1 - \langle\sigma_k\rangle_c^2$,
recovering the familiar statement that information gain purifies
fastest when the measured observable is least
certain~\cite{Jacobs2014_Book}.

\medskip
\noindent\textit{Remark.}
Equation~\eqref{eq:purity_mono} is the explicit derivative at a fixed
conditional state: the stationary expectation
$\mathbb{E}_\mathrm{ss}[\mathcal{P}]$ depends on $\eta$ also through
the trajectory measure itself, so Proposition~1 identifies the
dominant mechanism rather than supplying a closed monotonicity
theorem for $\mathbb{E}_\mathrm{ss}[\mathcal{P}]$. The ordering it
predicts is, however, borne out pointwise in our benchmarks: the
$\etaa = 0.35$ estimate has lower purity than the $\etatrue = 0.7$
truth on $99.9\%$ of trajectories (Table~\ref{tab:overcert}).

\subsection{The geometric loophole: from purity to coherence}
\label{sec:geometric_loophole}

Proposition~1 concerns the production of \emph{purity}
$\mathcal{P} = (1+|\mathbf{r}|^2)/2$, not of coherence
$S = \sqrt{|\mathbf{r}|^2 - z^2}$.
A shorter Bloch vector ($|\hat{\mathbf{r}}| < |\mathbf{r}|$) can
point closer to the equatorial plane (smaller $|\hat{z}|/|\hat{\mathbf{r}}|$),
yielding $\hat{S} > S$ even though
$\hat{\mathcal{P}} < \mathcal{P}$. We call this the
\emph{geometric loophole}.

Figure~\ref{fig:geometric} quantifies this loophole from $10^6$
Bloch-vector trajectories. The mechanism is traced to the
$\eta$-independent Rabi drive $\Ox\sigma_x$, which pushes the
estimated state equatorially just as hard as the true state, while the
weaker ($\propto\!\sqrt{\etaa}$) $\sigma_z$ backaction provides less
polar localization. On approximately $96\%$ of trajectories the
Bloch-vector length deficit dominates; on the remaining ${\sim}4\%$,
the angular shift wins.
The two decompositions describe the same events: the ${\sim}4\%$ of
trajectories on which the angular shift wins are precisely the
coherence-overcertification events $\hat{S} > S_\mathrm{true}$ of
Table~\ref{tab:overcert}.
Polar-angle histograms and exceedance probabilities behind these
aggregate numbers are shown in Supplementary Information Sec.~S13.

\subsection{Bias structure across estimators}

\begin{figure*}[t]
    \centering
    \includegraphics[width=\textwidth]{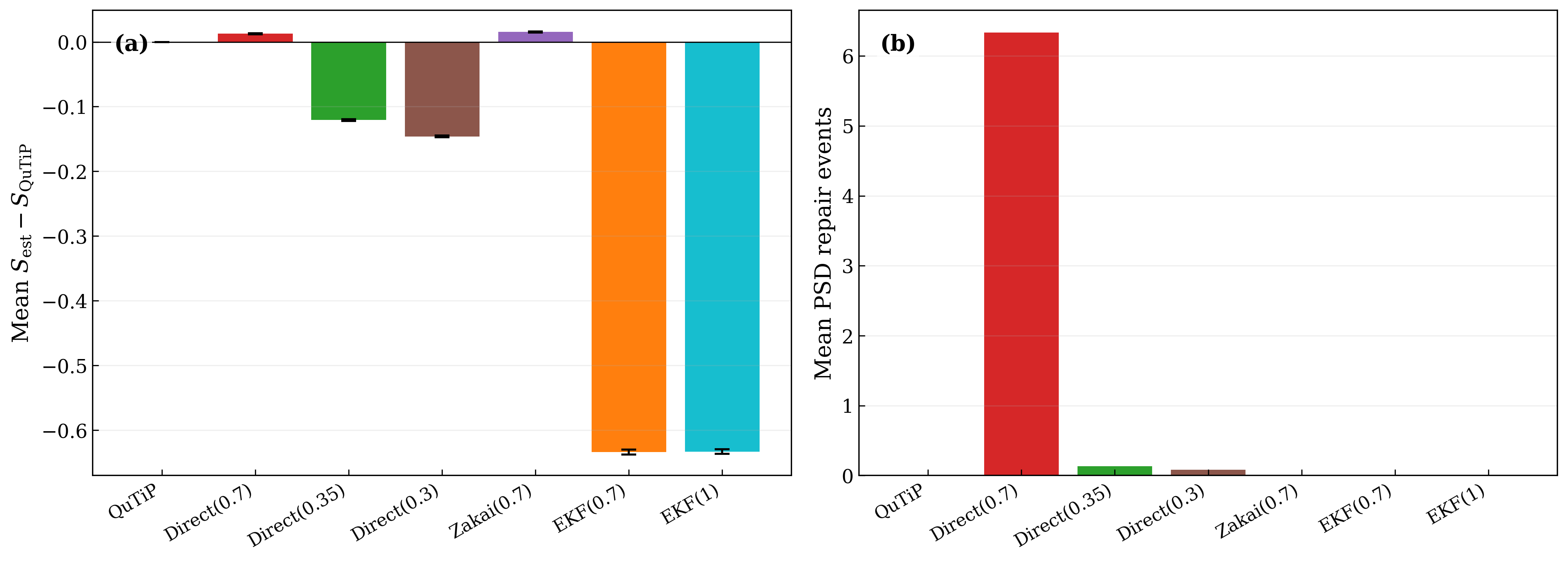}
    \caption{(a)~Mean signed coherence bias for each estimator.
    (b)~Mean PSD repair events per trajectory.}
    \label{fig:bias_repairs}
\end{figure*}

Figure~\ref{fig:bias_repairs} quantifies the bias structure across all
estimator configurations. Two distinct populations emerge. The
matched-efficiency Direct SME ($\etaa = 0.7$) shows near-zero average
bias but suffers ${\sim}7$ PSD-violating updates per
trajectory--numerical events where the integrator briefly produces
a non-physical state and must be repaired by projection back onto
the cone of positive-semidefinite operators.

 Each repair is a
discontinuous correction whose effect on the certified
coherence is hard to bound a priori, which is exactly the sort of
opacity that makes a security argument fragile. The conservative
Direct SME ($\etaa = 0.35$) inverts this picture: the bias becomes
modestly negative ($-0.13$), but PSD repairs drop to
${\lesssim}0.1$ per trajectory--two orders of magnitude lower.
The negative bias is a feature, not a bug. Because the QRNG
min-entropy bound and the entanglement-fidelity bound are both
\emph{lower} bounds on $S$, an estimator that systematically
under-reports $S$ produces conservative certificates: rejected
photons are sometimes coherent enough that they would have passed
under a perfect estimator, while the residual rate at which
accepted photons fall below the certified threshold is bounded
explicitly by the pointwise failure probability $\delta(\Sth, \varepsilon)$
developed in Sec.~\ref{sec:pointwise}.
The two EKF configurations show a much larger
conservative bias ($\approx -0.65$) that overcorrects to the point of
near-total rejection, which is why we discount them as practical
choices despite their numerical stability. The conservative Direct
SME thus occupies the operating sweet spot: bias small enough to
preserve useful heralding rates, repairs rare enough that the
trajectory dynamics are essentially analytic, and the residual error
characterized operationally by the OU bound of
Sec.~\ref{sec:pointwise} (with the structural exponential
supermartingale of Theorem~2 currently vacuous over the full
Bloch ball; see Sec.~\ref{sec:gap}).

\section{Pointwise overcertification bound}
\label{sec:pointwise}

Having established the average-case argument (Proposition~1) and
identified the geometric loophole (Sec.~\ref{sec:geometric_loophole}),
we now develop two complementary pointwise bounds on
overcertification. All pointwise simulations run in Bloch-vector
form rather than on the full density matrix--about $100\times$
faster, with consistency cross-checks against QuTiP given in
Supplementary Information Sec.~S10.

\subsection{The error process}

Both the true state $\mathbf{r}(t)$ and the estimate
$\hat{\mathbf{r}}(t)$ evolve under the SME in Bloch form,
driven by the same measurement record but with efficiencies $\etatrue$
and $\etaa$. The measurement backaction on Bloch
component $r_j$ from channel $k$ is
\begin{equation}
    \delta r_j = 2\sqrt{\eta\gamma_k}\,
    (\delta_{jk} - r_j r_k)\,dW_k,
    \label{eq:bloch_backaction}
\end{equation}
following from
$\mathrm{Tr}[\sigma_j\,\mathcal{H}[\sigma_k]\rho]
= 2(\delta_{jk} - r_j r_k)$.
(The deterministic drift contributions--Hamiltonian rotation,
dissipators, and the innovation-mismatch term defined below--are
standard and listed in Supplementary Information Sec.~S12.)
Define the squared-coherence error
\begin{equation}
    E(t) \equiv \hat{S}^2(t) - S^2(t)
    = (\hat{x}^2 + \hat{y}^2) - (x^2 + y^2).
    \label{eq:E_def}
\end{equation}
By It\=o's formula, $E(t)$ is a continuous semimartingale:
\begin{equation}
    dE = b(X)\,dt + \beta_x(X)\,dW^{(x)} + \beta_z(X)\,dW^{(z)},
    \label{eq:E_sde}
\end{equation}
where $X = (\mathbf{r}, \hat{\mathbf{r}}) \in K \subset \mathbb{R}^6$
with $K = \{|\mathbf{r}| \leq 1,\; |\hat{\mathbf{r}}| \leq 1\}$.
The drift contains three physically distinct pieces,
\begin{align}
    b(X) &= \underbrace{-2R_x(\hat{x}^2 - x^2)
    - 2R_y(\hat{y}^2 - y^2)
    - 2\Ox(\hat{y}\hat{z} - yz)}_{\text{restoring $+$ drive coupling}}
    \nonumber\\
    &\quad + \underbrace{4\,\etaa\, Q(\hat{\mathbf{r}})
    - 4\,\etatrue\, Q(\mathbf{r})}_{\text{It\=o correction}}
    \;+\; b_\mathrm{inn}(X),
    \label{eq:drift_decomp}
\end{align}
where $R_x = \grel/2 + 2\gdeph + 2\gz$ and
$R_y = \grel/2 + 2\gdeph + 2\gx + 2\gz$ are the transverse decay
rates,
\begin{equation}
    Q(\mathbf{r}) \equiv \gx\!\left[(1-x^2)^2 + x^2 y^2\right]
    + \gz\,(x^2 + y^2)\,z^2,
    \label{eq:Qdef}
\end{equation}
and $b_\mathrm{inn}(X)$ is the deterministic coupling generated by
expressing the estimator's innovation in terms of the true Wiener
increments,
$d\hat{W}_k = dW_k + 2\sqrt{\gamma_k}\,
\bigl(\sqrt{\etatrue}\,r_k - \sqrt{\etaa}\,\hat{r}_k\bigr)\,dt$;
its full expression, with the complete drift and diffusion
coefficients, is listed in Supplementary Information Sec.~S12. For
$\hat{\mathbf{r}} = \mathbf{r}$ the It\=o correction reduces to
$4(\etaa - \etatrue)\,Q(\mathbf{r}) \leq 0$: this conservative forcing
is what drives the stationary mean of $E$ below zero. The full drift
is, however, \emph{not} sign-definite on $K$
(dense sampling of the full joint Bloch ball exhibits drift values
$b \approx 2.0$ at corners the dynamics never visits, so
$\sup_K b > 0$, while the visited region reaches only ${\approx}0.8$;
Sec.~\ref{sec:gap})--a fact that separates the operational bound of
Sec.~\ref{sec:OU_bound}, fitted on the visited region, from the
structural bound of Theorem~2, which must pay for the global
supremum.

\subsection{Ornstein--Uhlenbeck comparison (operational bound)}
\label{sec:OU_bound}

The coupled measurement-and-drive dynamics generate a noisy but
approximately restoring evolution for the error process, which motivates
an Ornstein--Uhlenbeck (hereafter OU) approximation:
$dE \approx -\mu(E - \bar{E})\,dt + \sigma_E\,d\tilde{W}$,
with steady-state mean $\bar{E} = \nu/\mu < 0$.
We extract the OU parameters from
$10^6$ trajectories by conditional drift fitting
(bin by $E$, compute $\langle dE/dt \mid E \rangle$, fit the
linear relation). The same parameters can be recovered from the
steady-state distribution and from the autocorrelation function;
we cross-check both in Supplementary Information Sec.~S11.

\begin{figure*}[t]
    \centering
    \includegraphics[width=\textwidth]{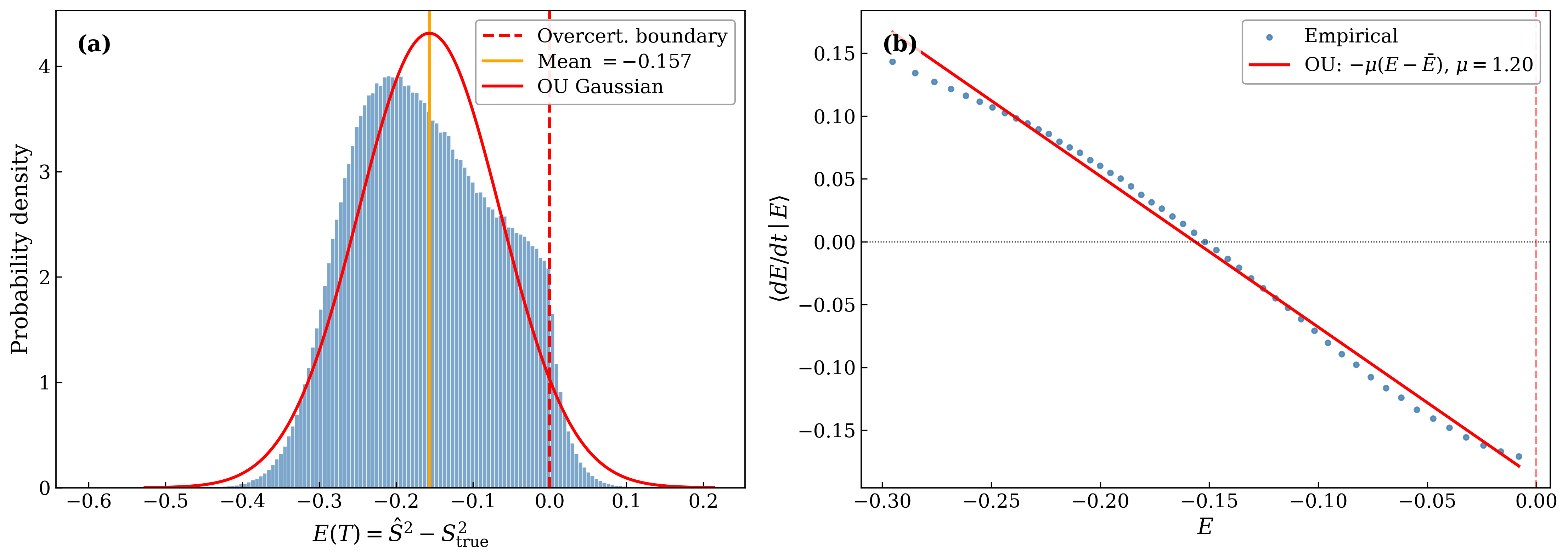}
    \caption{OU characterization from $10^6$ trajectories.
    (a)~Distribution of $E(T)$; the OU Gaussian (red) fits well.
    (b)~Conditional drift $\langle dE/dt \mid E \rangle$ is linear
    in $E$ (blue dots), with OU fit (red) giving $\mu = 1.20$,
    $\bar{E} = -0.157$.}
    \label{fig:pointwise_E}
\end{figure*}

Figure~\ref{fig:pointwise_E} validates the OU model.\footnote{The OU
parameters $\mu$, $\sigma_E$, and $\nu$ carry units of rate and are
quoted in the time units of the pointwise simulation, in which the
rates of Table~\ref{tab:params} enter without the explicit $2\pi$
factor (consistent with $\mu T \approx 12.0$ at $T = 10$). The
dimensionless error statistics $\bar{E}$ and $\mathrm{Var}(E)$--and
hence the overcertification rates and the failure probabilities
$\delta(\Sth, \varepsilon)$--are invariant under a global rescaling of all
rates and do not depend on this convention.}
The extracted parameters are:
\begin{equation}
    \mu = 1.20, \quad \bar{E} = -0.157, \quad
    \sigma_E = 0.142, \quad \nu = -0.188,
    \label{eq:OU_params}
\end{equation}
with steady-state variance $\mathrm{Var}(E) = 0.0086$ in good
agreement with the OU stationarity prediction
$\sigma_E^2/(2\mu) \approx 0.0084$. Because $\mu T \approx 12.0 \gg 1$
and $E(0) = 0$ (Sec.~\ref{sec:protocol}), the law of $E(T)$ is
stationary to excellent approximation--transient corrections to the
mean and variance are $O(e^{-\mu T}) \sim 10^{-5}$--so all tails
below are evaluated with the stationary Gaussian. The OU stationary
distribution gives:
\begin{equation}
    \Pr[\hat{S}(T) > S_\mathrm{true}(T)]
    \leq \tfrac{1}{2}\,\mathrm{erfc}\!\left(
    \frac{|\bar{E}|}{\sqrt{2\,\mathrm{Var}(E)}}\right)
    = 4.5\%.
    \label{eq:OU_bound}
\end{equation}
The empirical overcertification rate is $3.7\%$, confirming that the
OU bound is conservative (Table~\ref{tab:overcert}).

\medskip
\noindent
\textbf{From error tails to certification failure.}
The security statements of Sec.~\ref{sec:composable} concern not the
raw exceedance $\{\hat{S} > S + \varepsilon\}$ but the
\emph{acceptance--failure} event: the router accepts,
$\hat{S}(T) > \Sth$, while the true coherence falls below the
certified value, $S_\mathrm{true}(T) < \Sth - \varepsilon$. A
one-line geometric observation places this event much deeper in the
tail of $E$ than $\varepsilon^2$:

\medskip
\noindent\textbf{Lemma 1} (Acceptance--failure level).
\textit{If $\hat{S}(T) > \Sth$ and
$S_\mathrm{true}(T) < \Sth - \varepsilon$ for some
$0 \leq \varepsilon \leq \Sth$, then}
\begin{equation}
    E(T) \,>\, \Sth^2 - (\Sth - \varepsilon)^2
    \,=\, \varepsilon\,(2\Sth - \varepsilon).
    \label{eq:level}
\end{equation}
\noindent\textit{Proof.}
$E = \hat{S}^2 - S_\mathrm{true}^2
> \Sth^2 - (\Sth - \varepsilon)^2$. \hfill$\square$

\medskip
\noindent
Since $\varepsilon(2\Sth - \varepsilon) \geq \varepsilon^2$ whenever
$\Sth \geq \varepsilon$--satisfied at every operating point we
consider--the acceptance--failure event lies deeper in the $E$-tail
than the loose raw-exceedance implication $E > \varepsilon^2$, and
accordingly receives a smaller probability.
We therefore \emph{define} the certification failure probability as
the joint quantity
\begin{align}
    \delta(\Sth, \varepsilon)
    &\equiv \Pr\!\bigl[\hat{S}(T) > \Sth \,\wedge\,
    S_\mathrm{true}(T) < \Sth - \varepsilon\bigr] \nonumber\\
    &\leq \tfrac{1}{2}\,\mathrm{erfc}\!\left(
    \frac{\varepsilon\,(2\Sth - \varepsilon) + |\bar{E}|}
    {\sqrt{2\,\mathrm{Var}(E)}}\right),
    \label{eq:delta_def}
\end{align}
with the stationary OU parameters of Eq.~\eqref{eq:OU_params}.
Equation~\eqref{eq:delta_def} reduces to Eq.~\eqref{eq:OU_bound} at
$\varepsilon = 0$ and, unlike the raw tail, depends explicitly on
$\Sth$: falsifying a higher certified threshold requires a larger
estimation error, so $\delta$ improves with $\Sth$ at fixed
$\varepsilon$. By contrast, the raw event
$\{\hat{S} > S + \varepsilon\}$ implies only $E > \varepsilon^2$
(because $\hat{S} + S > \varepsilon$), and the corresponding OU
tails--$3.6\%$ at $\varepsilon = 0.1$ and $1.7\%$ at
$\varepsilon = 0.2$ (Table~\ref{tab:overcert})--are loose against
the empirical rates precisely because the implication discards the
typically order-unity factor $\hat{S} + S \approx 2\,S_\mathrm{typ}$.
No element of the security analysis uses these raw-event bounds.

\begin{table}[tb]
\centering
\caption{Overcertification rates from $10^6$ trajectories. The OU
column applies the stationary Gaussian tail at the level implied by
each event: $E > 0$ for the first row and $E > \varepsilon^2$ for the
raw exceedance rows (the implication
$\hat{S} > S + \varepsilon \Rightarrow E > \varepsilon^2$ discards
the factor $\hat{S} + S$, which is why these bounds are loose; the
security analysis instead uses the acceptance--failure level of
Lemma~1, cf.\ Table~\ref{tab:composable}).}
\label{tab:overcert}
\small
\begin{tabular}{lcc}
\toprule
Quantity & OU bound & Empirical \\
\midrule
$\Pr[\hat{S} > S_\mathrm{true}]$ & $4.5\%$ & $3.7\%$ \\
$\Pr[\hat{S} > S_\mathrm{true} + 0.1]$ & $3.6\%$ & $0.09\%$ \\
$\Pr[\hat{S} > S_\mathrm{true} + 0.2]$ & $1.7\%$ & $<0.001\%$ \\
\midrule
$\Pr[\hat{\mathcal{P}} > \mathcal{P}]$ (purity) & -- & $0.10\%$ \\
Geometric amplification & -- & ${\sim}40\times$ \\
\bottomrule
\end{tabular}
\end{table}

The OU comparison is operational rather than fully rigorous: it
relies on the empirical observation that $E(t)$ is well-approximated
by a Gaussian Ornstein--Uhlenbeck process with the parameters in
Eq.~\eqref{eq:OU_params}, validated on $10^6$ trajectories. A
strictly model-independent tail bound, valid without OU
approximation, follows from the exponential supermartingale developed
next; the gap between these two bounds is the subject of
Sec.~\ref{sec:gap}.

\subsection{Exponential supermartingale (structural bound)}

A model-independent bound follows from an exponential supermartingale.

\medskip
\noindent\textbf{Theorem 2} (Exponential overcertification bound).
\textit{Suppose both filters are initialized at the same state, so
$E_0 = 0$, and define}

\begin{equation}
\begin{aligned}
    c_\alpha &= \sup_{X \in K}\Lambda_\alpha(X), \\
    \Lambda_\alpha(X) &= \alpha\,b(X)
    + \tfrac{\alpha^2}{2}\bigl(\beta_x(X)^2 + \beta_z(X)^2\bigr).
\end{aligned}
    \label{eq:c_alpha}
\end{equation}
\textit{Then for every level $a > 0$ and horizon $T > 0$,}
\begin{equation}
    \Pr\!\left[\sup_{t \leq T} E_t \geq a\right]
    \leq \inf_{\alpha > 0}
    \exp\!\bigl(-\alpha a + T c_\alpha\bigr).
    \label{eq:sm_bound}
\end{equation}
\textit{In particular, $a = \varepsilon^2$ bounds the raw exceedance
$\Pr[\hat{S}_T \geq S_T + \varepsilon]$, and
$a = \varepsilon(2\Sth - \varepsilon)$ bounds the acceptance--failure
probability $\delta(\Sth, \varepsilon)$ of Eq.~\eqref{eq:delta_def}
(Lemma~1).}

\medskip
\noindent\textit{Proof.}
The coefficients $\beta_x, \beta_z$, and $b$ are polynomials in the
components of $X$, hence bounded on the compact set $K$, which is
forward-invariant for the dynamics. Define
$M_t^{(\alpha)} = \exp\bigl(\alpha E_t -
\int_0^t \Lambda_\alpha(X_s)\,ds\bigr)$. By It\=o's formula,
$dM^{(\alpha)} = M^{(\alpha)}\,\alpha\,
(\beta_x\,dW^{(x)} + \beta_z\,dW^{(z)})$, so $M^{(\alpha)}$ is a
positive local martingale; boundedness of $\beta_{x,z}$ on $K$
verifies Novikov's condition, making it a true martingale with
$M_0^{(\alpha)} = 1$ (this is where $E_0 = 0$ enters; an offset
$E_0 \neq 0$ would contribute a factor $e^{\alpha E_0}$). Let
$\tau_a = \inf\{t \leq T : E_t \geq a\} \wedge T$, a bounded stopping
time, so optional stopping gives
$\mathbb{E}[M_{\tau_a}^{(\alpha)}] = 1$. On the event
$\{\sup_{t \leq T} E_t \geq a\}$, the pointwise bound
$\Lambda_\alpha(X_s) \leq c_\alpha$ gives
$M_{\tau_a}^{(\alpha)} \geq e^{\alpha a - T c_\alpha}$, whence
Eq.~\eqref{eq:sm_bound} follows by Markov's inequality and
optimization over $\alpha > 0$. (Here and throughout, $T c_\alpha$ is
to be read as $T \max(c_\alpha, 0)$; in our setting
$\sup_K b > 0$ ensures $c_\alpha > 0$.)
\hfill$\square$

The drift, diffusion, and supremum quantities entering Theorem~2,
together with the numerical evaluation of $c_\alpha$ over the full
Bloch ball, are written out in Supplementary Information
Sec.~S12.

\subsection{The gap between operational and rigorous bounds}
\label{sec:gap}

\begin{figure}[tb]
    \centering
    \includegraphics[width=\columnwidth]{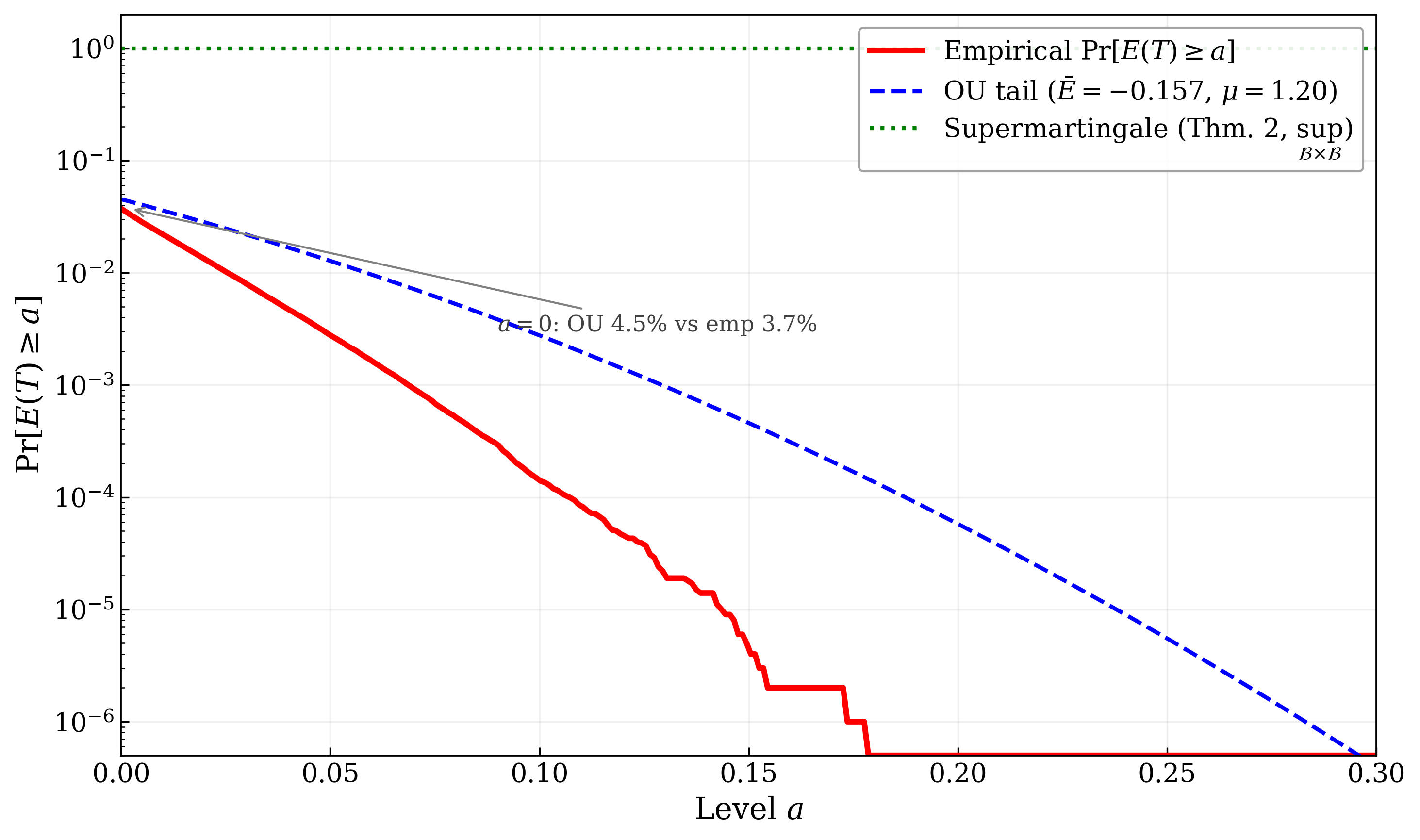}
    \caption{Tails of the error process at the decision time, as a
function of the level $a$. Red: empirical $\Pr[E(T) \geq a]$ from
$10^6$ trajectories. Blue dashed: stationary OU tail, conservative at
every level ($1.2\times$ safety factor at $a = 0$). Green dotted:
supermartingale full-ball diagnostic [Eq.~\eqref{eq:sm_bound}],
obtained by dense Monte-Carlo sampling of $\Lambda_\alpha$ over
$\mathcal{B}\times\mathcal{B}$ (a sampled maximum, not a certified
supremum); the sampled positive drift--$b \approx 2.0 > 0$ at
Bloch-ball corners the dynamics never visits--already renders it
vacuous ($\equiv 1$ at every level), and since a Monte-Carlo maximum
lower-bounds the true $\sup_K \Lambda_\alpha$, the rigorous global
bound is at least as vacuous.
Certification events enter this axis at the levels of Lemma~1:
$a = \varepsilon(2\Sth - \varepsilon)$ for
acceptance--failure, $a = \varepsilon^2$ for raw exceedance.}
    \label{fig:tail_bounds}
\end{figure}

Figure~\ref{fig:tail_bounds} compares the three tails as functions of
the error level $a$. The OU tail tracks the empirical one with a
$1.2\times$ safety factor at $a = 0$ and remains conservative at
every level. The
supermartingale bound is structural: Theorem~2 holds rigorously, but
the supremum $c_\alpha$ taken over the entire Bloch ball is
numerically vacuous because $b(X)$ attains positive values at corners
the dynamics never visits--dense sampling exhibits $b \approx 2.0$,
hence $\sup_K b > 0$--so the
right-hand side of Eq.~\eqref{eq:sm_bound} exceeds~1 across the
range of levels considered.

The theorem itself is mathematically rigorous; the gap is
computational, concentrated in $c_\alpha$--the supremum of a
degree-4 polynomial over a semialgebraic set. Two approaches can
tighten it: (i)~restricting $K$ to the forward-invariant reachable
set; or (ii)~sum-of-squares (SOS) polynomial
optimization~\cite{Parrilo2003}.
Until that gap is closed, the OU comparison
[Eq.~\eqref{eq:OU_bound}] supplies the operational guarantee on
$\delta(\Sth, \varepsilon)$, and the supermartingale of Theorem~2
contributes the structural existence of an exponential tail bound.

\subsection{Composable security statement}
\label{sec:composable}

Regardless of which bound supplies the failure probability
$\delta(\Sth, \varepsilon)$ [the OU tail of Eq.~\eqref{eq:delta_def}, or
Theorem~2 with $a = \varepsilon(2\Sth - \varepsilon)$ once the
$c_\alpha$ gap is closed], the composable structure is the same. Both
corollaries are stated as joint (``accept-and-fail'') probabilities,
the standard abort-tolerant form of composable security: no
conditioning on acceptance is performed, so no division by the
heralding rate appears. In smooth-entropy language,
$\delta(\Sth, \varepsilon)$ plays the role of the smoothing parameter:
outside a failure set of probability at most $\delta(\Sth, \varepsilon)$,
the accepted conditional state obeys the pointwise bound of
Eq.~\eqref{eq:Hmin_composable}.

\medskip
\noindent\textbf{Corollary 1} (Composable min-entropy).
\textit{Except with probability at most $\delta(\Sth, \varepsilon)$, a
routing event is either rejected or its true conditional coherence
satisfies $S_\mathrm{true}(T) \geq \Sth - \varepsilon$, in which case}
\begin{equation}
    H_\infty \geq -\log_2\!\left(
    \frac{1 + \sqrt{1 - (\Sth - \varepsilon)^2}}{2}\right).
    \label{eq:Hmin_composable}
\end{equation}
\textit{Over $n$ rounds with re-initialized qubits
(Sec.~\ref{sec:protocol}), the accepted events certify $H_\infty$
bits each, except with total probability at most
$n\,\delta(\Sth, \varepsilon)$.}

\medskip
\noindent\textbf{Corollary 2} (Composable entanglement fidelity).
\textit{For two-node entanglement distribution under assumptions
A1--A4, with the two nodes operating independently, except with
probability at most
$1 - (1 - \delta(\Sth, \varepsilon))^2 \leq 2\,\delta(\Sth, \varepsilon)$
a coincidence of accepted heralds certifies}
\begin{equation}
    \Fmm \geq \frac12
    + \frac{(\Sth - \varepsilon)^2}{2\,\bigl(2 - (\Sth-\varepsilon)^2\bigr)}.
    \label{eq:Fent_composable}
\end{equation}

\begin{figure*}[t]
    \centering
    \includegraphics[width=\textwidth]{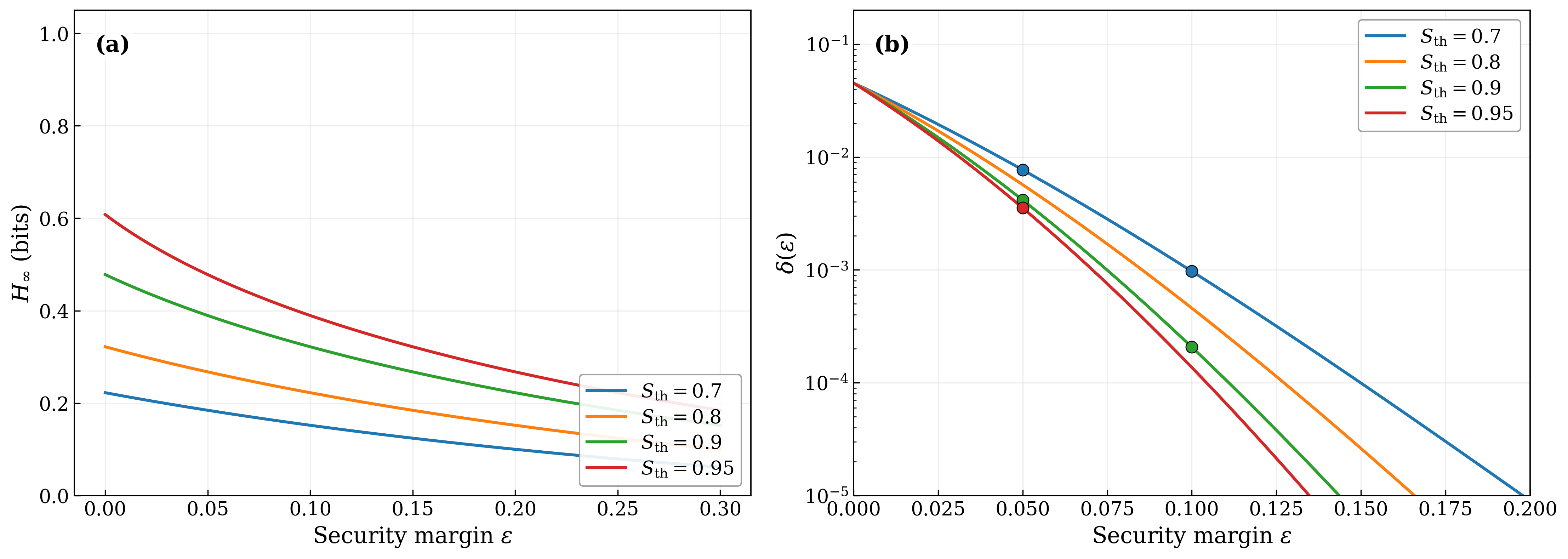}
    \caption{Composable security.
    (a)~Min-entropy $H_\infty$ vs.\ security margin $\varepsilon$ for
    various $\Sth$.
    (b)~OU joint acceptance--failure probability $\delta(\Sth, \varepsilon)$
    [Eq.~\eqref{eq:delta_def}], one curve per $\Sth$, evaluated at the
    Lemma-1 level $a = \varepsilon(2\Sth - \varepsilon)$; markers indicate
    the operating points of Table~\ref{tab:composable}.}
    \label{fig:composable}
\end{figure*}

Figure~\ref{fig:composable} traces these guarantees as functions of
the security margin $\varepsilon$: panel~(a) shows how the certified
min-entropy degrades as $\varepsilon$ is raised at fixed $\Sth$, and
panel~(b) gives the OU acceptance--failure probability
$\delta(\Sth, \varepsilon)$ [Eq.~\eqref{eq:delta_def}] that sets the
confidence with which each bound holds; with this joint
acceptance--failure definition, $\delta$ improves with $\Sth$ at
fixed $\varepsilon$, so the two knobs are mildly synergistic rather
than independent.
The two knobs play distinct roles. The threshold $\Sth$ trades
certified quality against \emph{heralding rate}: raising $\Sth$
delivers a higher $H_\infty$ and $\Fmm$ floor but accepts fewer
trajectories. The security margin $\varepsilon$ trades certified
quality against \emph{failure probability} at fixed $\Sth$:
tightening $\varepsilon$ (smaller $\varepsilon$) raises $\delta$,
loosening it (larger $\varepsilon$) lowers $\delta$ but lowers the
certified $H_\infty$ and $\Fmm$ as well. Heralding rate is set
purely by $\Sth$ and is independent of the choice of $\varepsilon$.
Representative operating
points are tabulated in Table~\ref{tab:composable}, and the sweep
is extended to additional $(\Sth, \varepsilon)$ pairs in Table~S3
(Supplementary Information Sec.~S14).

\begin{table}[tb]
\centering
\caption{Composable security operating points [OU bound,
Eq.~\eqref{eq:delta_def}]. $\delta(\Sth, \varepsilon)$ is the joint
acceptance--failure probability of Lemma~1; $H_\infty$ and $\Fmm$ are
the certified floors of Corollaries~1 and~2.}
\label{tab:composable}
\begin{tabular}{ccccc}
\toprule
$\Sth$ & $\varepsilon$ & $\delta(\Sth, \varepsilon)$ &
$H_\infty$ [bits] & $\Fmm$ \\
\midrule
0.70 & 0.05 & $0.8\%$  & 0.18 & 0.63 \\
0.70 & 0.10 & $0.1\%$  & 0.15 & 0.61 \\
0.90 & 0.05 & $0.4\%$  & 0.39 & 0.78 \\
0.90 & 0.10 & $0.02\%$ & 0.32 & 0.74 \\
0.95 & 0.05 & $0.4\%$  & 0.48 & 0.84 \\
\bottomrule
\end{tabular}
\end{table}

At $\Sth = 0.9$ and $\varepsilon = 0.05$, the protocol certifies
$H_\infty \geq 0.39$ bits per accepted photon except with failure
probability $0.4\%$ ($99.6\%$ confidence; Table~\ref{tab:composable}).

\section{Estimator hierarchy and numerical benchmarking}
\label{sec:estimators}

To evaluate the impact of estimator choice on certification quality, we
implement four estimator classes spanning seven configurations
(Table~\ref{tab:estimators}), each receiving the same QuTiP-generated
measurement records for fair comparison. Update equations for the
Zakai and Direct SME variants are listed in Supplementary
Information Secs.~S15 and~S16.

\begin{table}[tb]
\centering
\caption{Estimator configurations benchmarked.}
\label{tab:estimators}
\begin{tabular}{lcl}
\toprule
Label & Dynamics & $\etaa$ \\
\midrule
QuTiP SME   & Full (reference) & 0.7 \\
Direct SME  & Full, PSD repair & 0.7, 0.35, 0.3 \\
Zakai       & Linear, unnormalized & 0.7 \\
EKF         & Linear, Bloch-ball projection & 0.7, 1.0 \\
\bottomrule
\end{tabular}
\end{table}

\subsection{Results: estimator hierarchy}

\begin{figure*}[t]
    \centering
    \includegraphics[width=\textwidth]{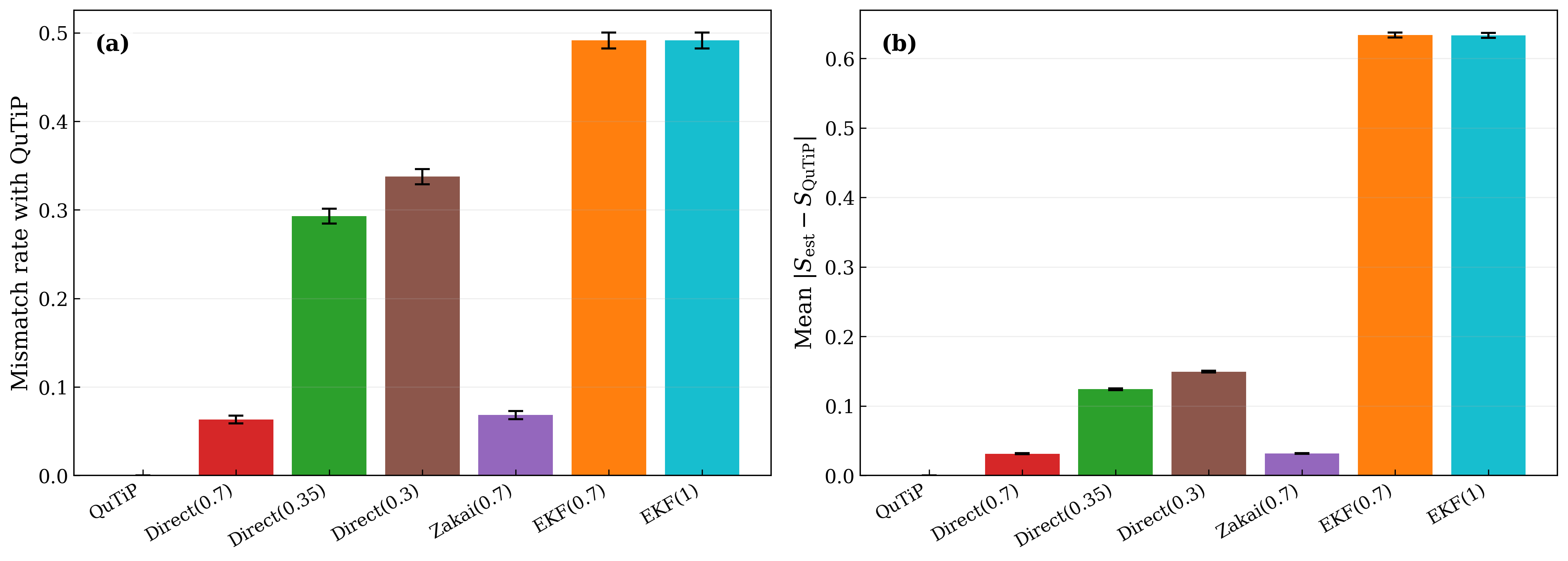}
    \caption{Estimator performance across 3000 trajectories.
    (a)~Decision mismatch rate at $\Sth = 0.7$.
    (b)~Mean absolute coherence estimation error.}
    \label{fig:performance}
\end{figure*}

\begin{figure}[tb]
    \centering
    \includegraphics[width=\columnwidth]{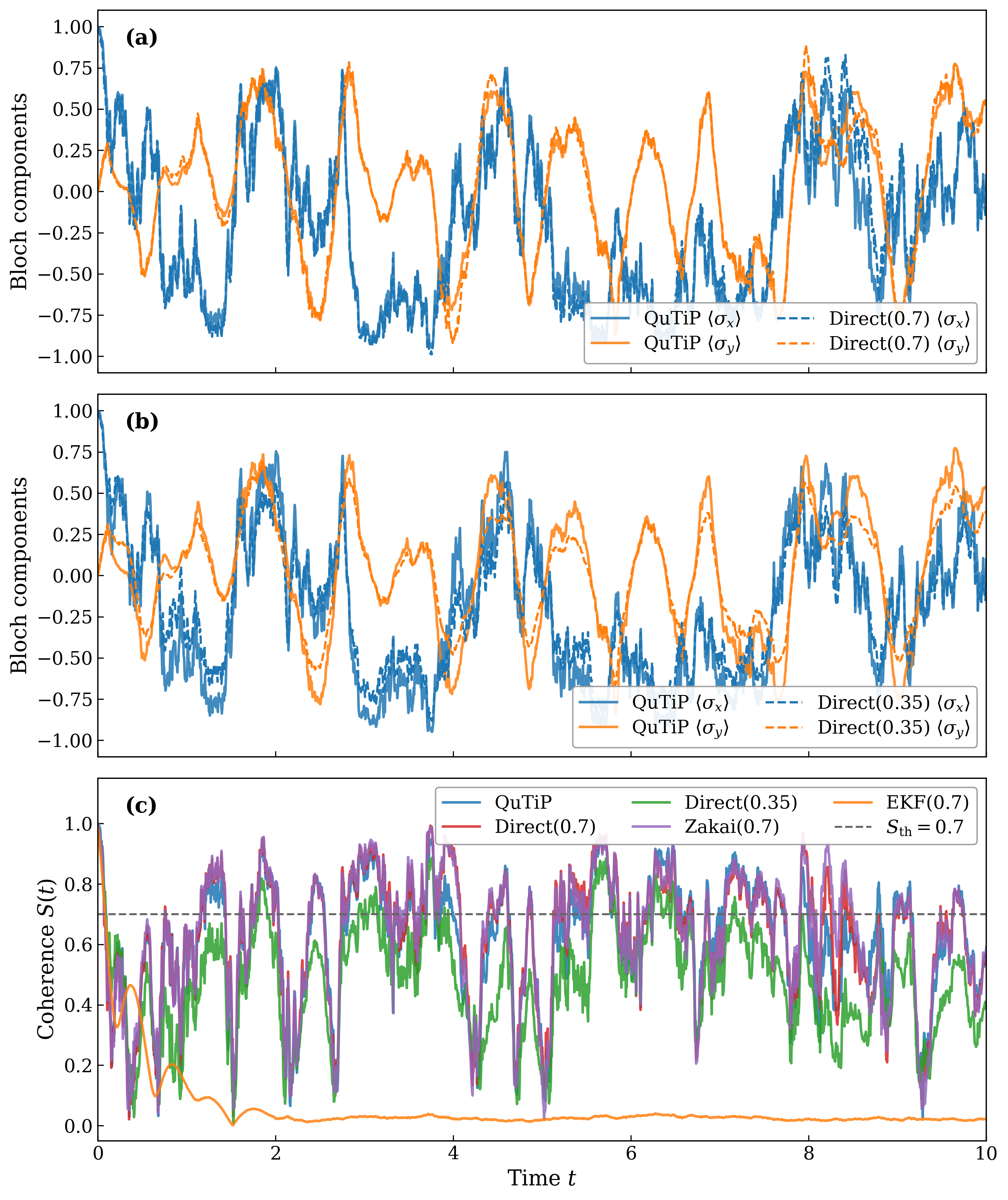}
    \caption{Single-trajectory comparison.
    (a)~Direct SME ($\etaa = 0.7$) vs.\ QuTiP.
    (b)~Direct SME ($\etaa = 0.35$) vs.\ QuTiP.
    (c)~Coherence $S(t)$ from all estimator classes.}
    \label{fig:trajectory}
\end{figure}

\begin{figure*}[t]
    \centering
    \includegraphics[width=\textwidth]{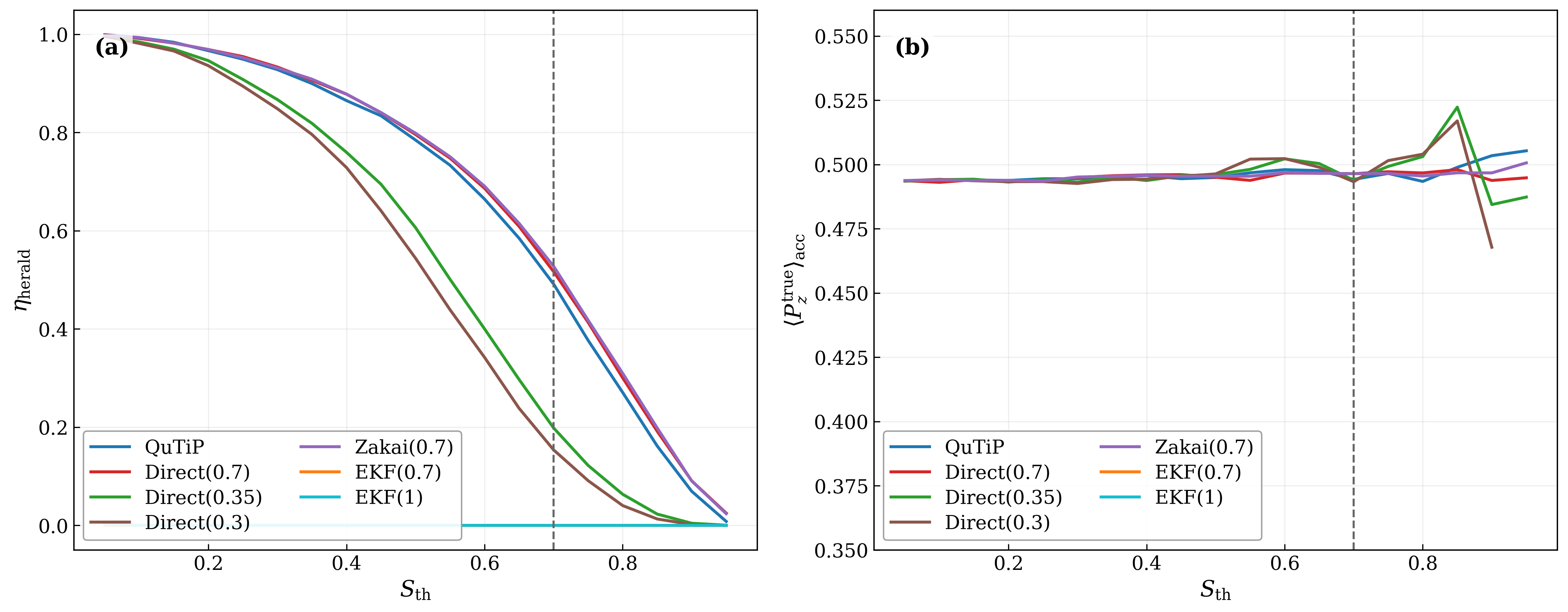}
    \caption{Threshold sweep across $\Sth \in [0.05, 0.95]$.
    (a)~Heralding efficiency vs.\ $\Sth$.
    (b)~Accepted-event mean true Born-rule probability
    $\langle P_z^\mathrm{true}\rangle_\mathrm{acc}$ vs.\ $\Sth$.}
    \label{fig:threshold}
\end{figure*}

Figures~\ref{fig:performance}--\ref{fig:threshold} summarize the
estimator hierarchy. 

The mismatch and bias statistics in
Fig.~\ref{fig:performance} aggregate over 3000 trajectories, the
single-trajectory comparison in Fig.~\ref{fig:trajectory} shows the
qualitative behavior on representative runs, and Fig.~\ref{fig:threshold}
sweeps the routing threshold to expose the heralding-efficiency trade.

Each estimator class exhibits a distinct failure mode:
\begin{itemize}
    \item \textbf{Direct SME (matched $\etaa = 0.7$):}
    Low-bias and low-mismatch, but ${\sim}7$ PSD violations per
    trajectory. Numerically unsafe.
    \item \textbf{Direct SME (conservative $\etaa = 0.35$, $0.3$):}
    Negative bias ($-0.13$ to $-0.16$), ${\lesssim}0.1$ PSD repairs.
    Safe for all downstream applications; preferred for FPGA deployment.
    \item \textbf{Zakai:} Low bias and mismatch, zero PSD repairs,
    but weight drift at long times.
    \item \textbf{EKF:} ${\sim}{-}0.65$ coherence bias at both
    $\etaa = 0.7$ and $\etaa = 1.0$; the insensitivity to $\etaa$
    (Fig.~\ref{fig:performance}) indicates the bias originates in the
    linearization itself rather than in the efficiency assumption.
    Unsuitable.
\end{itemize}

The conservative Direct SME ($\etaa = 0.35$) emerges as the
recommended solution: numerically stable, safely biased, and
hardware-efficient at ${\sim}300$ operations per step.
Three further diagnostics support this picture: per-estimator
coherence-score distributions (Fig.~S1; Supplementary Information
Sec.~S3), confusion matrices at $\Sth = 0.7$ (Fig.~S3; Sec.~S5), and
precision--recall curves (Fig.~S4; Sec.~S6).

\section{Assumed efficiency: decision accuracy versus overcertification}
\label{sec:analytical}

Both the EKF and the Direct SME are run with $\etaa < \etatrue$, and it
is worth being precise about why. One might expect the assumed
efficiency to admit an \emph{optimal} interior value that minimizes the
estimator's decision mismatch against the QuTiP reference. A
systematic sweep of the mismatch versus $r = \etaa/\etatrue$ across
six parameter sets (Supplementary Information Secs.~S7--S8,
Figs.~S5--S6) shows that it does not: in every set the decision
mismatch decreases monotonically as $r \to 1$, with the empirical
optimum at $r^*_\mathrm{emp} \approx 0.8$--$1.0$, i.e.\ at or near
\emph{matched} efficiency. This is what one expects of a
maximum-likelihood filter--supplying the true efficiency makes the
running estimate track the conditional state most faithfully. It also
overturns a heuristic balance argument between measurement-induced
estimation-error accumulation and relaxation that would place the
optimum near $r \approx 0.4$--$0.5$ via a scaling
$\etaa^*/\etatrue \approx 1/(1+\xi)$ with
$\xi = 4\gamma S_\mathrm{typ}^2/R_b$; that scaling is \emph{not}
supported by the regenerated sweeps and is presented, as a transparent
negative result, in Supplementary Information Secs.~S7--S8.

The reason to nonetheless operate at $\etaa < \etatrue$ is not decision
accuracy but \emph{overcertification suppression}--the one-sided,
security-relevant criterion of
Secs.~\ref{sec:certification}--\ref{sec:pointwise}. Decision mismatch
penalizes both error directions symmetrically (rejecting a coherent
trajectory and accepting an incoherent one), whereas certification
integrity penalizes only the second. The matched-efficiency estimator,
which minimizes mismatch, carries a near-zero (slightly positive)
coherence bias and a correspondingly high false-acceptance rate at the
operating threshold ($133$ of $1526$ reject-class trajectories at
$\Sth = 0.7$; Supplementary Information Sec.~S5); underestimating $\eta$
injects a protective \emph{negative} bias [Fig.~\ref{fig:bias_repairs}]
that collapses that false-acceptance rate (to $0$ of $1526$ at
$\etaa = 0.35$) while also suppressing PSD-repair events by two orders
of magnitude (Sec.~\ref{sec:estimators}). The assumed efficiency is
therefore best read as a tunable \emph{safety margin}: lowering $\etaa$
buys a larger conservative bias--and hence a smaller overcertification
probability $\delta(\Sth, \varepsilon)$ [Eq.~\eqref{eq:delta_def}]--at the
cost of rejecting more marginal trajectories, with the operating point
set by the composable security budget of Sec.~\ref{sec:composable}
rather than by any universal ratio. For the deployed configuration we
adopt $\etaa = 0.35$, which sits well inside the conservative regime
(near-zero overcertification, ${\lesssim}0.1$ PSD repairs per
trajectory) while retaining a useful heralding rate.

\section{Discussion}
\label{sec:discussion}

\subsection{Hardware scaling: the role of \texorpdfstring{$T_1$}{T1}}

The parameter regime explored in this work
($\gamma/\gamma_\mathrm{dec} = 10$) is comfortably accessible in
present-day circuit-QED hardware, but state-of-the-art transmon
qubits push this ratio considerably further. Coherence times of
$T_1 \sim 300$--$1000\;\mu$s have been reported in tantalum-based
transmons~\cite{Wang2022_Transmon,Place2021_Tantalum} and
fluxonium~\cite{Somoroff2023}, with recent demonstrations approaching
the millisecond range~\cite{Tuokkola2025}. At a fixed measurement
rate $\gamma \sim 0.1\;$MHz, this translates into a
measurement-to-decoherence ratio of ${\sim}50\times$, where the
typical coherence on individual trajectories saturates near the
information-theoretic ceiling
$S_\mathrm{typ} \to \sqrt{2/3} \approx 0.82$. In this regime, the
heralding efficiency at $\Sth = 0.9$ rises substantially above the
${\sim}7\%$ figure of Table~\ref{tab:entanglement}, because a larger
fraction of the trajectory ensemble passes the threshold. The
practical implication is that the rate--quality tradeoff curve
shifts upward with each generation of qubit hardware: the same
device, run with the same software, produces more certified events
per second simply because the underlying conditional dynamics retain
coherence longer. This positions the protocol to benefit
automatically from ongoing hardware progress rather than requiring
fundamental redesign.

\subsection{Experimental feasibility}

The individual ingredients have been demonstrated in current
circuit-QED hardware: two-tone dispersive monitoring of $\sigma_x$
and $\sigma_z$ on a single transmon~\cite{HacohenGourgy2016_Simultaneous},
real-time FPGA-based Bayesian state
estimation~\cite{Vijay2012_Stabilizing,Ficheux2018}, tunable
couplers for conditional photon release~\cite{Chen2014_TunableCoupler},
and controlled qubit--cavity entangling interactions for
qubit--photon Bell-pair generation~\cite{Kurpiers2018,Zhong2021}.
Integrating all four in a single device
has not yet been demonstrated and represents the principal
experimental challenge. The key hurdle is achieving $\eta \gtrsim 0.5$
on both measurement channels simultaneously while preserving the
coherence budget required for $\Sth \geq 0.7$ at realistic
$T \sim \mu$s decision times. The conservative Direct SME requires
${\sim}300$ multiply-accumulate operations per update, fitting
within the per-step budget at practical FPGA update rates of
${\sim}100$--$200$\,ns. None of these requirements is
beyond the reach of a single dedicated experiment; what is
needed is the coordination engineering to bring them together,
which we view as a near-term goal rather than a long-term
aspiration.

\subsection{Trajectory-conditioned quantum devices}

Coherence-gated routing is, to our knowledge, the first instance of a
broader class of \emph{trajectory-conditioned quantum devices}: hardware
whose operation is gated by a real-time estimate of a continuous
quantum-information quantity, rather than by the projective collapse
of a single observable. The classification matters because the design
challenges are largely shared across the class. Any
trajectory-conditioned device must reconstruct a conditional state in
real time, must select a function of that state to threshold on, must
benchmark estimators against a ground-truth solver, and must bound the
probability that the estimate exceeds the truth. The methodology
developed here addresses all four: the SME-based estimator hierarchy
of Sec.~\ref{sec:estimators}, the certification-integrity analysis of
Sec.~\ref{sec:certification}, the OU comparison and supermartingale
structure of Sec.~\ref{sec:pointwise}, and the composable security
statements that follow.

Several other gating quantities suggest themselves. A
\emph{purity-gated} device would route on $\mathrm{Tr}(\rho_c^2)$
itself rather than its equatorial projection, with the application
target being heralded production of high-purity rather than
high-coherence states. An \emph{entropy-gated} device, conditioned
on the estimated von Neumann entropy of the partial trace of a
two-qubit conditional state, would herald entangled pairs of
guaranteed mixedness budget. A \emph{discord-gated} device, gating
on quantum discord between two continuously monitored qubits, would
provide a primitive for quantum-correlation-assisted protocols that
do not require entanglement. In each case, the analytic
structure--purity monotonicity, geometric loophole, OU
approximation, supermartingale bound--would need to be redeveloped
for the specific quantity, but the architectural template and the
estimator-as-security-primitive perspective transfer essentially
unchanged. We expect this template to be the more durable
contribution of the present work, with coherence-gated routing
serving as the worked example.

\section{Conclusions}
\label{sec:conclusions}

We have introduced coherence-gated routing: a single-photon routing
primitive conditioned on the real-time quantum coherence of a driven
qubit. The measurement basis, coherence basis, and Hamiltonian drive
axis form a self-consistent triad natural for circuit-QED hardware.

The central physical phenomenon--that individual quantum trajectories
retain large coherence while the ensemble average collapses to near
zero--was demonstrated directly (Fig.~\ref{fig:uncond_vs_cond}) and
provides the foundation for all subsequent results.

The coherence certification enables two primary applications:
(i)~a coherence-certified QRNG with min-entropy
bound $H_\infty \geq -\log_2\!\bigl(\frac{1+\sqrt{1-\Sth^2}}{2}\bigr)$,
and (ii)~a phase-tracked photon source for quantum networks.
Independent coherence certification at two nodes constrains the
matter--matter entanglement fidelity achievable via BSM to
$\Fmm \geq \tfrac12 + \Sth^2/[2\,(2-\Sth^2)]$ under idealized
assumptions; the product form $[(1+\Sth)/2]^2$ is \emph{not} a valid
lower bound for $\Sth \gtrsim 0.53$, because the uncertified
longitudinal tilt $z_i$ survives the feedforward and is correlated
across the two nodes by the BSM.

Through systematic estimator benchmarking, the conservative Direct SME
($\etaa < \etatrue$) emerged as the recommended solution: numerically
stable, safely biased, and hardware-efficient. Proposition~1
identifies the mechanism behind its safety--purity
production monotone in the assumed efficiency--an ordering realized
on $99.9\%$ of trajectories.

We developed two complementary pointwise overcertification bounds.
The Ornstein--Uhlenbeck comparison provides an operational bound of
$4.5\%$ on the raw overcertification probability, validated against
an empirical rate of $3.7\%$ from $10^6$ trajectories; through the
acceptance--failure level of Lemma~1 it yields sub-percent failure
probabilities for the composable security statements
(Corollaries~1 and~2). The exponential
supermartingale (Theorem~2) establishes the structural existence of
an exponential tail bound, with the computational gap concentrated
in a single polynomial optimization over the reachable state space.
We identified the geometric mechanism underlying the ${\sim}40\times$
amplification (at our operating point) from purity to coherence
overcertification: the
$\eta$-independent Rabi drive makes the estimated state slightly
more equatorial, partially compensating its shorter Bloch vector.
Closing the gap between the OU and supermartingale
bounds--via reachable-set restriction or sum-of-squares
optimization over the forward-invariant reachable set--is the
remaining step toward a fully rigorous composable security
statement; the OU-based bound already provides an operational
guarantee for near-term implementations. The broader
connection between numerical regularization and application-level
security applies to the class of trajectory-conditioned quantum
devices introduced here.


\section*{Data and code availability}
The simulation data and code that support the findings of this study are
available via Zenodo at https://doi.org/10.5281/zenodo.20799833.

\section*{Acknowledgements}
P.S.\ acknowledges fruitful discussions with Dr.\ Vasilii Vadimov and
Dr.\ Ujjwal Pratap.

\section*{Funding}
This study received no funding.

\section*{Author contributions}
P.S.\ conceived the project, developed the theoretical framework,
performed the simulations and analysis, and wrote the manuscript.

\section*{Competing interests}
The author declares no competing interests.

\section*{Additional information}
\textbf{Supplementary information} accompanies this paper. It contains the idealized
entanglement-distribution assumptions A1--A4 (Sec.~S1); the
reconditioning mechanism for conditional coherence (Sec.~S2);
coherence-score distributions, heralding scatter plots, confusion
matrices, and precision--recall curves for the estimator
benchmarking (Secs.~S3--S6); decision-mismatch sweeps across
parameter sets and the assessment of the $1/(1+\xi)$ scaling
(Secs.~S7--S8); the min-entropy
certification table (Sec.~S9); the pointwise simulation
methodology, OU parameter extraction, and supermartingale
$c_\alpha$ computation (Secs.~S10--S12); the quantitative
breakdown of the geometric loophole (Sec.~S13); extended composable
security operating points (Sec.~S14); and the Zakai filter and Direct
SME update equations (Secs.~S15--S16).

\textbf{Correspondence} and requests relating to this study should be addressed
to P.S.

  
\bibliographystyle{apsrev4-2}
\bibliography{bibliography}

\end{document}